\documentstyle[12pt,epsfig]{article}
\newcommand{\be}{\begin{equation}}
\newcommand{\ee}{\end{equation}}
\newcommand{\beq}{\begin{equation}}
\newcommand{\eeq}{\end{equation}}
\newcommand{\ba}{\begin{eqnarray}}
\newcommand{\ea}{\end{eqnarray}}
\newcommand{\bea}{\begin{eqnarray}}
\newcommand{\eea}{\end{eqnarray}}

\begin{document}
\baselineskip=15.5pt \pagestyle{plain} \setcounter{page}{1}
\begin{titlepage}

\leftline{OUTP-09 31 P} 

\vskip 0.8cm

\begin{center}

{\LARGE Holographic current correlators at finite coupling and
scattering off a supersymmetric plasma} \vskip .3cm

\vskip 1.cm

{\large {Babiker Hassanain\footnote{\tt babiker@thphys.ox.ac.uk}$
^{, \, a}$ and Martin Schvellinger\footnote{\tt
martin@fisica.unlp.edu.ar}$^{, \, b}$} }

\vskip 1.cm

{\it $^a$ The Rudolf Peierls Centre for Theoretical Physics, \\
Department of Physics, University of Oxford. \\ 1 Keble Road,
Oxford, OX1 3NP, UK. \\
$\&$ Christ Church College, Oxford, OX1 1DP, UK.} \\

\vskip 0.5 cm

{\it $^b$ IFLP-CCT-La Plata, CONICET and \\
Departamento  de F\'{\i}sica, Universidad Nacional de La Plata.
\\ Calle 49 y 115, C.C. 67, (1900) La Plata,  \\ Buenos Aires,
Argentina.} \\

\vspace{1.cm}

{\bf Abstract}

\end{center}

By studying the effect of the ${\cal {O}}(\alpha'^3)$ string
theory corrections to type IIB supergravity, including those
corrections involving the Ramond-Ramond five-form field strength, we
obtain the corrected equations of motion of an Abelian perturbation
of the AdS$_5$-Schwarzschild black hole. We then use the gauge
theory/string theory duality to examine the coupling-constant
dependence of vector current correlators associated to a gauged
$U(1)$ sub-group of the global ${\cal {R}}$-symmetry group of
strongly-coupled ${\cal {N}}=4$ supersymmetric Yang-Mills theory at
finite temperature. The corrections induce a set of
higher-derivative operators for the $U(1)$ gauge field, but their
effect is highly suppressed. We thus find that the ${\cal
{O}}(\alpha'^3)$ corrections affect the vector correlators only
indirectly, through the corrected metric. We apply our results to
investigate scattering off a supersymmetric Yang-Mills plasma at low
and high energy. In the latter regime, where Deep Inelastic
Scattering is expected to occur, we find an enhancement of the
plasma structure functions in comparison with the infinite 't Hooft coupling result.

\noindent

\end{titlepage}

\newpage

\tableofcontents

\newpage
\vfill

\section{Introduction}

The gauge theory/gravity duality
\cite{Maldacena:1997re,Gubser:1998bc, Witten:1998qj,Aharony:1999ti}
is a powerful tool that may hold the key to unlocking the mysteries
of strongly-coupled field theory. Although the applicability of the
duality has so far been limited to theories with a high degree of
symmetry, it is nonetheless important to understand the
strong-coupling regime of these theories fully, as they share many
features with QCD. A very important and phenomenologically relevant
example of a system which can be studied using the AdS/CFT
correspondence is the quark-gluon plasma (QGP). This system, which
can be obtained as a result of the collision of two heavy nuclei,
such as those collisions achieved at RHIC, has shown tantalizing
hints of ideal hydrodynamic behaviour (for some review articles see
\cite{Shuryak:2003xe,Gyulassy:2004zy,Muller:2007rs,
CasalderreySolana:2007zz,Shuryak:2008eq,Heinz:2008tv,Iancu:2008sp}).

On the other hand, if one applies the gauge theory/string theory
duality to calculate the hydrodynamic properties of the large $N$
limit of an ${\cal {N}}=4$ supersymmetric Yang-Mills plasma with
gauge group $SU(N)$, one finds results that may be close to those
measured for the QGP \cite{Shuryak:2003xe}. A lot of attention has
therefore been devoted to the study of the various transport
coefficients of supersymmetric plasma in the hydrodynamic regime, in
an effort to more accurately model the behaviour of the quark-gluon
plasma studied at RHIC. This programme of approaching the real-world
QGP using AdS/CFT techniques requires taking into account many
non-trivial aspects, such as the fact that the QGP contains
fundamental quarks, whereas ${\cal{N}}=4$ supersymmetric Yang-Mills
theory only has fields in the adjoint representation of the gauge
group. Another major ingredient that must be included is going
beyond the infinite 't Hooft parameter limit, as the QGP is governed
by large yet finite coupling. On the supergravity side, this
corresponds to adding string-theoretic higher-curvature corrections
to the gravitational background.

The hydrodynamic regime describes the behaviour of plasma at
distance scales $ \gg 1/T$, where $T$ is the temperature of the
plasma. One may choose to study plasma in the opposite
regime, namely for distance scales $\ll 1/T$, by using a different
physical probe. One can deduce many useful
properties of plasma by considering its response to a hard probe,
such as a photon or a parton. In this work we are concerned with
Deep Inelastic Scattering (DIS) in strongly-coupled ${\cal {N}}=4$
supersymmetric plasma at finite 't Hooft coupling. As is well-known,
DIS is a key process for investigating hadronic structure. A lepton
scatters from a hadron through the emission of a virtual photon with
four-momentum $q^\mu$. This photon then probes the hadron structure
at distances of order $\sqrt{1/q^2}$, thus giving us valuable
information about the distribution of momenta amongst the various
partons comprising the hadron. The relevant quantity for studying
deep inelastic scattering is the matrix element of two
electromagnetic currents inside the hadron. In particular, the
hadronic tensor is defined in terms of structure functions which can
be extracted from the imaginary part of the two-point function of
the electromagnetic current.

There are two distinct regimes for this process, separated by the
coupling strength of the theory. At weak coupling the appropriate
description is given by perturbative QCD. The operator product
expansion of the current-current correlator is dominated by
twist-two operators in the perturbative regime. On the other hand, at strong coupling, the
operator product expansion is dominated by double trace operators,
and the process can be studied using the gauge/string theory duality
\cite{Polchinski:2002jw}.

The situation for plasma is different but entirely analogous to that
of scattering off a single hadron. One may now view the scattering
as taking place off ``quasiparticles'' which constitute the plasma.
Clearly, a knowledge of the structure functions governing this
process yields valuable information on the dynamics of plasma in an
important regime which is not probed in the hydrodynamic limit. In
particular, deep inelastic scattering in ${\cal {N}}=4$ SYM plasma
at infinite 't Hooft coupling has been reported in several recent
articles \cite{Hatta:2007he,Hatta:2007cs,Hatta:2008tx,Iancu:2008sn,
Hatta:2009ra,Avsar:2009xf,Bayona:2009qe,Dominguez:2009cm,Iancu:2009py}.
In the holographic dual description, finite 't Hooft coupling corrections to the structure functions of
an ${\cal {N}}=4$ SYM plasma can be investigated by considering the
effect of higher-curvature terms on the vector fluctuations of the
metric. This is the subject of the present article.

So far, the community has focussed a lot of its effort on finite 't Hooft coupling
corrections to the transport properties of plasma in the
hydrodynamic regime (see for instance
\cite{Buchel:2004di,Benincasa:2005qc,Janik:2006gp,Armesto:2006zv,Buchel:2008sh,Myers:2008yi,Buchel:2008ae,
Ritz:2008kh,Myers:2009ij,Cremonini:2009sy}). Such transport properties include the
shear viscosity and the mass-density diffusion constants, both of
which can be obtained by studying tensor fluctuations of the
supergravity metric with higher-curvature corrections. One the other hand,
the vector fluctuations of the metric yield quantities such as the charge
diffusion and conductivity. The finite coupling corrections to these
quantities have been considered so far only for the cases where the
additional curvature terms have been of mass-dimension four and six.
The case of ${\cal{N}}=4$ supersymmetric Yang-Mills plasma, where
the stringy corrections are known, and are found to yield dimension
eight operators, has not been considered. Therefore, the formalism
we develop below for computing the effect of these dimension eight
corrections on the vector fluctuations, can be applied to the computation of the charge conductivity and
diffusion constants for ${\cal{N}}=4$ supersymmetric Yang-Mills
plasma with 't Hooft coupling corrections. In this paper, we are not
concerned with the hydrodynamic regime, choosing instead to focus on
the effect of the finite coupling corrections on the structure
functions of the plasma. We will consider the effect of the
corrections on the charge transport properties in a future work \cite{us-hydrodynamics}.

Let us define the premise of the paper more carefully. We
investigate the full effect of the ${\cal {O}}(\alpha'^3)$ string theory corrections to
the retarded correlators of the vector currents associated with a
gauged $U(1)$ sub-group of the global ${\cal {R}}$-symmetry group of
${\cal {N}}=4$ supersymmetric YM theory. This allows us to compute
the structure functions $F_s$ with $s=1, 2$ of a strongly-coupled ${\cal {N}}=4$
supersymmetric YM plasma with gauge group $SU(N)$ for finite values
of the 't Hooft coupling. The structure functions are extracted from
the imaginary part of the retarded current-current commutator
\be
R_{\mu\nu}(q) = i \int d^4x \, e^{-i q \cdot x} \, \Theta(x_0) \,
<[J_\mu(x), J_\nu(0)]> \, ,
\ee
where $\Theta(x_0)$ is the Heaviside function, while $J_\mu(x)$ is
the conserved current associated with the gauged $U(1)$ subgroup
mentioned above. The expectation value is understood as a thermal
average over the statistical ensemble of an ${\cal {N}}=4$ SYM
plasma at temperature $T$. It is assumed that in this plasma the
tensor $R_{\mu\nu}(q)$ plays an analogous role as the hadronic
tensor does in deep inelastic scattering off a single hadron. In
that case, the imaginary part of the hadronic tensor allows us to
extract the structure functions of the hadron, which in perturbative
QCD describe the partonic nature of the hadron.

The tensor structure of the retarded current-current commutator can
be derived from two properties:  $J_\mu(x)$-current conservation and
$R_{\mu\nu}(q) = R_{\nu\mu}(-q)$, so that
\be
R_{\mu\nu}(q) = \left(\eta_{\mu\nu} - \frac{q_\mu q_\nu}{Q^2}
\right) \, R_1 + \left[ n_\mu \, n_\nu - \frac{n \cdot q}{Q^2}
(n_\mu q_\nu + n_\nu q_\mu) + \frac{q_\mu q_\nu}{(Q^2)^2} (n \cdot
q)^2 \right] \, R_2 \, ,
\ee
where the flat four-metric $\eta_{\mu\nu}$ is chosen with mostly
plus signature $(-1, 1, 1, 1)$, while $n^\mu$ is the plasma
four-velocity and $Q^2$ is the virtuality, defined as $Q^2 =
q^2-\omega^2$. In the plasma rest frame $n^\mu=(1, 0, 0, 0)$.
We have also defined $q^\mu=(\omega, 0, 0, q)$ as the momentum transfer.
Thus, $q \cdot n=-\omega$ and this is a negative quantity.

We define the DIS plasma structure functions as follows
\be
F_1(x_B, Q^2) \equiv \frac{1}{2 \pi} \, \textrm{Im} R_1(x_B, Q^2) \,,
\ee
and
\be
F_2(x_B, Q^2) \equiv \frac{-(n \cdot q)}{2 \pi T} \, \textrm{Im}
R_2(x_B,Q^2) \, .
\ee
Notice that we have defined a new Bjorken variable which involves
the temperature
\be
x_B = - \frac{Q^2}{2 (q \cdot n) T} = \frac{Q^2}{2 \omega T} \, .
\ee
This paper has two main results: the first is that the 't Hooft coupling
corrections to the retarded current-current correlators for the electromagnetic $U(1)$ group
enhance the plasma structure functions. This is the physical result of the
formalism developed below, as far as DIS is concerned.

On a more formal level, at distance scales $\ll 1/T$, the computations reported here imply that
the higher-curvature ${\cal {O}}(\alpha'^3)$ operators affect the on-shell action of the vector perturbations only through the correction
to the metric. This applies to the full set of corrections involving
the metric and the Ramond-Ramond five-form field strength appearing
in type IIB supergravity. We show below that the corrections to the
Maxwell equations due to these terms only appear at very high powers
of the radial coordinate $u$, so that their effect vanishes at the
boundary $u\to 0$. 

The paper is organized as follows. In section 2 we briefly describe
the basic action and metric setup for the computation of the
retarded current-current correlators at infinite 't Hooft coupling.
In section 3 we study the string theory corrections to the metric,
introducing also the {\it ansatz} for the vector fluctuations of the
ten-dimensional metric and the Ramond-Ramond five-form field
strength. In section 4 we perform a detailed study of the
contributions of the type IIB string theory action at ${\cal
{O}}(\alpha'^3)$ which contain the Ramond-Ramond five-form field
strength. In section 5 we derive the Maxwell equations for vector
fluctuations with the ${\cal {O}}(\alpha'^3)$ corrected metric. We
then consider deep inelastic scattering and explicitly show the
enhancement of the longitudinal and transverse plasma structure
functions due to the mentioned string theory corrections. In the
conclusions presented in section 6 we discuss our results. The full
expression of the $C^4$ term when vector fluctuations are included
is presented in Appendix A. In Appendix B we provide a detailed
analysis of the equations of motion, along with the subtleties
arising from the higher derivative terms introduced by the
higher-curvature corrections to the gauge field action. In Appendix
C we give details of how to solve the equations of motion with
finite 't Hooft coupling corrections for the low energy regime, the physical
interpretation of which is a multiple scattering series.

\section{Infinite 't Hooft coupling}

At infinite 't Hooft coupling, the string theory holographic dual to
finite-temperature ${\cal {N}}=4$ SYM theory is the
AdS-Schwarzschild black hole solution with a five-sphere as the
internal space. This is a solution of type IIB supergravity with
only the leading curvature terms, namely the Einstein-Hilbert action
coupled to the dilaton and the five-form field strength:
\be
S_{10}=\frac{1}{2 \kappa_{10}^2}\int \, d^{10}x\,
\sqrt{-G}\left[R_{10}-\frac{1}{2}\left(\partial\phi
\right)^2-\frac{1}{4.5!}\left(F_5\right)^2 \right] \,
.\label{action-10D}
\ee
It is easy to check that the solution of this system for a constant
dilaton and $N$ units of five-form flux through the five-sphere is
given by the metric
\be
ds^2 = \frac{(\pi T R)^2}{u} \, (-f(u) dt^2 +d\vec{x}^2) +
\frac{R^2}{4 u^2 f(u)} du^2 + R^2 d\Omega_5^2 \, ,
\label{uncorrected}
\ee
where $f(u)=1-u^2$ and $R$ is the radius of the AdS$_5$ and the
five-sphere. In these coordinates the AdS-boundary is at $u=0$ while
the black hole horizon is at $u=1$. We denote the AdS$_5$
coordinates by the indices $m$, where $m= \{(\mu=0, 1, 2, 3),5\}$.

The AdS/CFT correspondence stipulates that supergravity fields are
dual to certain field-theory operators. In this work we are
concerned with the conserved current $J_\mu(x)$ associated with a
gauged $U(1)$ subgroup of the $SU(4)$ ${\cal{R}}$-symmetry group
possessed by the ${\cal {N}}=4$ SYM theory \cite{CaronHuot:2006te}.
The supergravity field corresponding to $J_\mu(x)$ is the $s$-wave
(massless) mode of the vector fluctuation about the background of
Eq.(\ref{uncorrected}). More precisely, we introduce off-diagonal
fluctuations $G_{\mu a}$ of the metric, where $a$ is an index on the
five-sphere, and plug this fluctuating metric into the
ten-dimensional action Eq.(\ref{action-10D}), making sure that we
are picking a specific Abelian subgroup
\cite{Argurio:1998cp,Cvetic:1999xp}. We employ the following {\it
ansatz} for the perturbed metric \cite{Cvetic:1999xp,Chamblin,Tran},
for a general case which includes the black brane metric with
obvious substitutions, where we have imposed that the internal
metric is the five-sphere
\ba
ds^2&=&\left[ g_{mn}+ \frac{4}{3} R^2 \, A_m A_n \right] \,dx^m
dx^n + R^2 \, d\Omega_5^2 \nonumber \\
&& + \frac{4}{\sqrt{3}} R^2 \left(\sin^2y_1 \, dy_3 + \cos^2 y_1 \,
\sin^2 y_2 \, dy_4 + \cos^2 y_1 \, \cos^2 y_2 \, dy_5 \right) \, A_m
\, dx^m \, \label{metric-ansatz}.
\ea
We write the metric of the unit five-sphere as $d\Omega_5^2$ where
\be
d\Omega_5^2  = dy_1^2 + \cos^2 y_1 \, dy_2^2  + \sin^2 y_1 \, dy_3^2
+ \cos^2 y_1 \, \sin^2 y_2 \, dy_4^2  + \cos^2 y_1 \ \cos^2y_2 \,
dy_5^2 \, .\nonumber
\ee
Also, we use the reduction {\it ansatz} for the five-form field
strength $F_5 = G_5 + \ast G_5$, where:
\be
G_5 = -\frac{4}{R} \epsilon_5 + \frac{R^3}{\sqrt{3}} \, \left(
\sum_{i=1}^3 d\mu_i^2 \wedge d\phi_i \right) \wedge \overline{\ast}
F_2 \, , \label{F5ansatz}
\ee
while $F_2 = dA$ is the Abelian field strength and $\epsilon_5$ is
the volume form of the five-dimensional metric of the
AdS-Schwarzschild black hole. The Hodge dual $\ast$ is taken with
respect to the ten-dimensional metric, while $\overline{\ast}$
denotes the Hodge dual with respect to the 5D metric piece of the
black hole. In addition
\ba
&& \mu_1 = \sin y_1 \, ,\,\,\,\,\,\,\,\,\,\,\,\,\,\,\,\,\,\,\, \mu_2
= \cos y_1 \, \sin y_2 \, , \,\,\,\,\,\,\,\,\,\,\,\,\,\,\,\,\,\,\,
\mu_3 = \cos y_1 \, \cos y_2 \, , \\
&& \phi_1 = y_3 \, ,
\,\,\,\,\,\,\,\,\,\,\,\,\,\,\,\,\,\,\,\,\,\,\,\,\, \phi_2 = y_4 \, ,
\,
\,\,\,\,\,\,\,\,\,\,\,\,\,\,\,\,\,\,\,\,\,\,\,\,\,\,\,\,\,\,\,\,\,\,\,\,\,\,\,\,\,\,
\phi_3 = y_5 \, .
\ea
Inserting the above {\it ans\"atze} for the type IIB supergravity
fields into the zeroth-order supergravity action of
Eq.(\ref{action-10D}) and discarding all the higher (massive)
Kaluza-Klein harmonics of the five-sphere, we are then left with the
following action for the zero-mode Abelian gauge field $A_m$:
\be
S = -\frac{N^2}{64 \pi^2 R} \int d^4x \, du \, \sqrt{-g} \, g^{mp}
\, g^{nq} \, F_{mn} \, F_{pq} \, , \label{action}
\ee
where the Abelian field strength is $F_{mn}=\partial_m A_n -
\partial_n A_m$, the partial derivatives are $\partial_m =
\partial/\partial x^m$, while $x^m=(t, \vec{x}, u)$, with $t$ and
$\vec{x}=(x_1, x_2, x_3)$ refer to the Minkowski coordinates, and $g
\equiv \textrm{det} (g_{mn})$, where the latter is the metric of
AdS-Schwarzschild black hole.

The equations of motion derived from the above action are just the
Maxwell equations for the bulk five-dimensional Abelian gauge fields
$A_m$ on the AdS-Schwarzschild spacetime Eq.(\ref{uncorrected}). By
studying the bulk solutions of these equations subject to certain
boundary conditions that we will specify shortly, we can obtain the
retarded correlation functions \cite{Son:2002sd, Policastro:2002se}
of the operator $J_\mu(x)$. At the level of this section, these
correlators would pertain to the infinite 't Hooft coupling limit. Our aim is
to obtain the leading coupling-constant dependence of these
correlators. We now describe how this is achieved.

\section{The ${\cal {O}}(\alpha'^3)$ string theory corrections}
\label{corrections-section}

We would like to derive the leading order $\alpha'$-corrected action
for the vector fluctuations of the metric. The higher-curvature
corrections on the supergravity side correspond to finite-coupling
corrections in the field theory. In other words, for any given
field-theoretic observable $\mathcal{O}$, we can write a series
${\mathcal{O}}_0+{\mathcal{O}}_1/\lambda^{n_1}+\cdots$, where
$\lambda$ is the 't Hooft coupling, and $n_1$ is a positive number
which indicates that the lowest order correction to the result at
infinite coupling ${\mathcal{O}}_0$ need not begin at order one. The
inclusion of higher-derivative corrections to the supergravity must
take place at the level of the ten-dimensional action, through the
evaluation of stringy corrections to Eq.(\ref{action-10D}). The
leading corrections were found to begin at
${\mathcal{O}}(\alpha'^3)$. There is a large volume of literature on
these corrections, and the initial application to holography was at
zero temperature \cite{Banks:1998nr}, where the metric was found to
remain AdS$_5 \times S^5$, verifying certain non-renormalization
theorems of CFT correlators. At finite temperature
\cite{Gubser:1998nz,Pawelczyk:1998pb}, much of the work focussed on
the corrections to the thermodynamics of the black hole. The
corrections were then revisited in references
\cite{deHaro:2002vk,deHaro:2003zd,Peeters:2003pv}, where the
computation of the $\alpha'$-corrected metric was improved and
attempts were made to address the issue of the completeness of the
corrections at leading order in $\alpha'$. More recently, the
conjectured lower bound for the ratio of the shear viscosity to the
entropy of any material
\cite{Policastro:2001yc,Kovtun:2004de}\footnote{More recent studies
have shown that the conjectured universal lower bound does not hold
when certain higher-derivative corrections are included. For a
discussion see \cite{Hofman:2009ug,Sinha:2009ev} and references
therein.}, has prompted interest in the higher curvature
corrections to the supergravity duals of gauge theories, primarily
in the spin-2 sector of the fluctuations \cite{Buchel:2004di,Buchel:2008sh}. In
\cite{Paulos:2008tn,Myers:2008yi} the higher curvature corrections
to the dual of  ${\cal {N}}=4$ SYM were parsed thoroughly to
determine how they affect the metric. Our case is slightly more
complicated, because we must use the corrected metric as our
background and must also evaluate the action for the vector
fluctuations of the metric, thereby obtaining the corrected
Lagrangian for the field $A_\mu$. There are therefore two distinct
parts to the calculation: the first part consists of obtaining the
minimal gauge-field kinetic term using new perturbed {\emph{and}}
corrected metric and five-form {\it ans${\ddot{a}}$tze}. The second
part of the computation consists of obtaining the corrections to the
gauge field Lagrangian coming directly from the higher-derivative
operators. The reason why these two steps are distinct is that the
first step will require insertion of the corrected perturbation {\it ans${\ddot{a}}$tze} into the minimal 10D supergravity two-derivative part
Eq.(\ref{action-10D}). The second step requires insertion of the
{\it{uncorrected}} perturbation {\it ans${\ddot{a}}$tze} into the
higher-curvature terms in ten dimensions.

The corrections to the 10D action are given by \cite{Myers:2008yi}
\be\label{10DWeyl}
S_{10}^{\alpha'}=\frac{R^6}{2 \kappa_{10}^2}\int \, d^{10}x\,
\sqrt{-G}\left[ \, \gamma e^{-\frac{3}{2}\phi}W_4 + \cdots\right] \, ,
\ee
where $\gamma$ encodes the dependence on the 't Hooft coupling
$\lambda$ through the definition $\gamma \equiv \frac{1}{8} \,
\xi(3) \, (\alpha'/R^2)^{3}$, with $R^4 = 4 \pi g_s N \alpha'^2$.
Setting $\lambda = g_{YM}^2 N \equiv 4 \pi g_s N$, $\gamma$ becomes
\be
\gamma \equiv   \frac{1}{8} \, \xi(3) \, \frac{1}{\lambda^{3/2}} \,
.
\ee
The $W_4$ term is a dimension-eight operator, and is given by
\be
W_4=C^{hmnk} \, C_{pmnq} \, C_h^{\,\,\,rsp} \, C^{q}_{\,\,\,rsk} +
\frac{1}{2} \, C^{hkmn} \, C_{pqmn} \, C_h^{\,\,\, rsp} \,
C^q_{\,\,\, rsk} \, ,
\ee
where $C^{q}_{\,\,\, rsk}$ is the Weyl tensor. The dots in
Eq.(\ref{10DWeyl}) denote extra corrections containing contractions
of the five-form field strength $F_5$, which we can schematically
write as $\gamma
(C^3{\mathcal{T}}+C^2{\mathcal{T}}^2+C{\mathcal{T}}^3+{\mathcal{T}}^4)$,
where $C$ is the Weyl tensor and ${\mathcal{T}}$ is a tensor found
in \cite{Myers:2008yi} and composed of certain combinations of
$F_5$. The authors of \cite{Myers:2008yi} showed that the metric
itself is only corrected by $W_4$, essentially due to the vanishing
of the tensor ${\mathcal{T}}$ on the uncorrected supergravity solution.
After taking into account the contribution of this term to the Einstein
equations, one finds the corrected metric
\cite{Gubser:1998nz,Pawelczyk:1998pb,deHaro:2003zd}
\be
ds^2 = \left(\frac{r_0}{R}\right)^2\frac{1}{u} \, \left(-f(u) \,
K^2(u) \, dt^2 + d\vec{x}^2\right) + \frac{R^2}{4 u^2 f(u)} \,
P^2(u) \, du^2 + R^2 L^2(u) \, d\Omega_5^2 \, ,\label{proper-metric}
\ee
where
\ba
K(u) &=& \exp{[\gamma \, (a(u) + 4b(u))]} \, , \\
P(u) &=& \exp{[\gamma \, b(u)]} \, ,\\
L(u) &=&  \exp{[\gamma \, c(u)]} \, .
\ea
and
\ba
a(u) &=& -\frac{1625}{8} \, u^2 - 175 \, u^4 + \frac{10005}{16} \, u^6 \, , \\
b(u) &=& \frac{325}{8} \, u^2 + \frac{1075}{32} \, u^4 - \frac{4835}{32} \, u^6 \, , \\
c(u) &=& \frac{15}{32} \, (1+u^2) \, u^4 \, .
\ea
Notice that $r_0$ is related to the temperature by
\be
r_0 = \frac{\pi T R^2}{(1+\frac{265}{16} \gamma)} \, ,
\ee
so that there is a hidden but important $\gamma$-dependence inside
$r_0$. The reader should be aware that there is some confusion in
the literature regarding the $\alpha'$-corrected metric, and we
refer the reader to \cite{Pawelczyk:1998pb} for a discussion. We use
the metric of \cite{Pawelczyk:1998pb}, and our conventions follow
theirs' closely, with the obvious change of coordinate
$u=r_0^2/r^2$.

Now that we know the corrected metric, we are able to obtain the
minimal kinetic term of the $U(1)$ gauge field. To this end, we must
construct the corrected versions of Eq.(\ref{metric-ansatz}) and
Eq.(\ref{F5ansatz}) and insert them into the two-derivative
supergravity action Eq.(\ref{action-10D}). The metric {\it ansatz}
we use is that of Eq.(\ref{metric-ansatz}) with the appropriate
corrected substitutions, and the imposition $R\to R \, L(u)$ to take
account of the non-factorisability of the corrected metric. As for
the {\it ansatz} for $F_5$ we use the fact that we are only
interested in the terms which are quadratic in the gauge-field
perturbations in order to define the following {\it ansatz}, which
is a direct extension of the unperturbed {\it ansatz} of Eq.(\ref{F5ansatz})
\be
G_5 = -\frac{4}{R} \overline{\epsilon} + \frac{R^3 L(u)^3}{\sqrt{3}}
\, \left( \sum_{i=1}^3 d\mu_i^2 \wedge d\phi_i \right) \wedge
\overline{\ast} F_2 \, . \label{F5ansatzCorrected}
\ee
Note that we are not interested in the part of $G_5$ which does not
contain the vector perturbations. This part is denoted by
$\overline{\epsilon}$, and only contributes to the potential of the
metric, and is thus accounted for by the use of the corrected metric
in the computation. Therefore, the only difference between this {\it
ansatz} and the uncorrected one as far as the gauge field is
concerned is the warp factor $L(u)$, which starts at
${\cal{O}}(u^4)$ and will be seen to drop out of all of our results.
Inserting the metric {\it ansatz} and the $F_5$ {\it ansatz} into
the action Eq.(\ref{action-10D}), we obtain the kinetic term for the
gauge field $A_m$, as expected:
\be
S = -\frac{N^2}{64 \pi^2 R} \int d^4x \, du \, \sqrt{-g} \, L^7(u)
\, g^{mp} \, g^{nq} \, F_{mn} \, F_{pq} \, , \label{Fsquared}
\ee
where the dependence on the dimensionless factor $L(u)$ is acquired
by the proper reduction from ten dimensions \cite{Kovtun:2003wp},
and ultimately arises as a consequence of the non-factorisability of
the corrected metric \cite{Pawelczyk:1998pb}. The determinant factor
$\sqrt{-g}$ refers to the five-dimensional part of the 10D metric of
Eq.(\ref{metric-ansatz}), and all 5D indices are raised and lowered
by that metric.

We have thus completed  the first step in our programme, that of
obtaining the minimal gauge kinetic term from the two-derivative
supergravity action. The next step is to obtain the effect of the
eight-derivative corrections of Eq.(\ref{10DWeyl}). Concretely, we
must determine the five-dimensional operators that arise once the
perturbed metric and five-form field strength {\it ans${\ddot{a}}$tze} are
inserted into Eq.(\ref{10DWeyl}). Crucially, we are able to use the
uncorrected {\it ans${\ddot{a}}$tze} Eq.(\ref{metric-ansatz}) and
Eq.(\ref{F5ansatz}) in this step, because using the corrected ones
results in terms of even higher order in $\gamma$. The salient point
to take from the discussion in the next section is that a simple
operator analysis together with an analysis of the equations of
motion reveals that the contributions arising directly from the
ten-dimensional higher-curvature operators {\emph{will not}
contribute to the on-shell action in this work.
We explain this important statement in the next section, where we
also introduce the explicit expressions for the ten-dimensional
eight-derivative corrections.

\section{The higher-curvature operators}
\label{10Dcorrections}

We first introduce the explicit expressions of the eight-derivative
corrections. In addition to $W_4=C^4$, we have the terms that we
denoted above by
$\gamma(C^3{\mathcal{T}}+C^2{\mathcal{T}}^2+C{\mathcal{T}}^3+{\mathcal{T}}^4)$.
The tensor $C$ is the Weyl tensor, and it only depends on the
metric. The six-tensor ${\mathcal{T}}$ is defined in terms of the
self-dual field $F^+= (1 + \ast) F_5/2$ via
\be
{\mathcal{T}}_{ABCDEF}=i\nabla_A F^+_{BCDEF}+\frac{1}{16}
\left[F^+_{ABCMN}F^+_{DEF}{}^{MN}-3 F^+_{ABFMN}F^+_{DEC}{}^{MN} \right] \, ,
\ee
where there is implicit antisymmetry in $[A,B,C]$ and $[D,E,F]$ in
addition to symmetry under the interchange $ [A,B,C]\leftrightarrow
[D,E,F]$. Note that for the purposes of this section we write the
ten-dimensional indices in capital letters, reserving small caps for
the AdS-Schwarzschild coordinates, and denoting the coordinates of
the five-sphere by indices with a tilde $\tilde{a}$. The six-tensor
${\mathcal{T}}$ is a complicated object in terms of its index
structure, but it is a rather simple object when viewed from the
point-of-view of the 5D effective field theory obtained upon
integrating out the sphere. The sheer size of the six-tensor
$\mathcal{T}$ means that it is very difficult (impossible) to
compute its contribution to the 5D gauge field Lagrangian directly.
We will therefore adopt a different approach below. We are
interested in evaluating $C_{ABCD}$ and ${\mathcal{T}}_{ABCDEF}$ on
the perturbed {\it ans${\ddot{a}}$tze} Eq.(\ref{F5ansatz}) and
Eq.(\ref{metric-ansatz}). We write the two tensors $C_{ABCD}$ and
${\mathcal{T}}_{ABCDEF}$ as
\ba
C_{ABCD}&=&C^{(0)}_{ABCD}+C^{(1)}_{ABCD}+C^{(2)}_{ABCD} \, , \nonumber \\
{\mathcal{T}}_{ABCDEF}&=&{\mathcal{T}}^{(0)}_{ABCDEF}+
{\mathcal{T}}^{(1)}_{ABCDEF}+{\mathcal{T}}^{(2)}_{ABCDEF} \, ,
\ea
where the superscript $(i)$ on each term in the right hand side of
this equation denotes the power of $F_{ab}$ contained within that
term, where $F_{ab}$ is the field strength of the $U(1)$ gauge
field. A crucial property of the six-tensor ${\mathcal{T}}_{ABCDEF}$
is that it vanishes when evaluated on the uncorrected $F_5$ and
$G_{MN}$ with no vectorial perturbations. In other words,
${\mathcal{T}}^{(0)}$ is zero. Given that we are only interested in
operators which are quadratic in $F_{ab}$, we can then completely
discard the operators $C{\mathcal{T}}^3+{\mathcal{T}}^4$, which is a
massive simplification. We then focus on the terms given by
$\gamma(C^4+C^3{\mathcal{T}}+C^2{\mathcal{T}}^2)$. We first note
that $C^4$ only contains the metric and no factors of the $F_5$ {\it
ansatz}. Our strategy will be to discuss the terms
$C^3{\mathcal{T}}+C^2{\mathcal{T}}^2$ and draw general conclusions
about the 5D operators that can arise from them. We will then
compute $C^4$ explicitly using the metric {\it ansatz}, and show
that it confirms our conclusions about the expected class of
operators. This is not a surprising outcome in a certain sense: the
terms containing ${\mathcal{T}}$ are nothing but the supersymmetric
completion of the $C^4$ term. Our final conclusion will be that
these operators all come in with a very high power in $u$, the
radial coordinate. Given that the holographic partition function is
evaluated in the ultraviolet, {\it i.e.} in the limit $u \to 0$, we
can prove that the only contributing operator is the minimal kinetic
term of Eq.(\ref{Fsquared}).

The important point to remember in the following analysis is that we
use the {\emph{uncorrected}} metric and five-form {\it ans${\ddot{a}}$tze}.
This has crucial consequences, the first being that the Weyl tensor
factorizes on the unperturbed metric because the latter is a direct
product. Moreover, the five-sphere has vanishing Weyl tensor. In
addition, the tensor ${\mathcal{T}}_{MNPQRS}$ is such that terms with one AdS
index and five internal $S^5$ indices are zero, and terms with one
internal index and five AdS indices are also zero. Finally, the fact
that we are only interested in terms that have two factors of $F_{i
j}$, where the latter is the $U(1)$ gauge field strength, simplifies
the analysis considerably. The final deduction that we require below
is that the terms coming from ${\mathcal{T}}^{(1,2)}$ will give rise
to five-dimensional operators composed solely of $g_{a b}$,
$F_{ij}$, $\nabla_k F_{ij}$ and the five-dimensional Levi-Civita
tensor $\epsilon_{abcde}$, where all the indices are
AdS-Schwarzschild indices. This is a direct consequence of the {\it
ansatz} for ${\mathcal{T}}^{(1,2)}$: at most it has two-derivatives,
and at least one of them must reside in the field strength $F_{ij}$.
Therefore, we cannot obtain terms that go like $R_{ij}$, the Ricci
tensor of the AdS-Schwarschild space, or like $R_{ijkl}$, because
both of these require two derivatives acting on the metric
components. This will become clear when we discuss the details of
the analysis.

\subsection{$C^2 {\mathcal{T}}^2$ terms}

In this section we will examine the ten-dimensional eight-derivative
terms and determine the five-dimensional gauge-field and gravity
operators that will result upon dimensional reduction. The approach
is based on a counting of the derivative terms and symmetries of
each particular term. Let us begin with the $C^2{\mathcal{T}}^2$
term. According to \cite{Paulos:2008tn} there are eight of these
terms, with different Lorentz contractions. They are given by:
\ba
C^2 {\mathcal{T}}^2&=&C_{ABCD} \, C_{ABCE} \, {\mathcal{T}}_{DGFHIJ} \,
{\mathcal{T}}_{EFGHIJ}  \nonumber \\
&&+\left(C_{ABCD} \, C_{ABEF} \, {\mathcal{T}}_{CDGHIJ} \, {\mathcal{T}}_{EFGHIJ} \,
+\, C_{ABCD} \, C_{AECF} \, {\mathcal{T}}_{BEGHIJ} \, {\mathcal{T}}_{DFGHIJ} \right. \nonumber \\
&&\left.+\, C_{ABCD} \, C_{AECF} \, {\mathcal{T}}_{BGHDIJ} \, {\mathcal{T}}_{EGHFIJ} \right) \nonumber \\
&&+\left[C_{ABCD} \, C_{AEFG} \, {\mathcal{T}}_{BCEHIJ} \, {\mathcal{T}}_{DFHGIJ} \,
+\, C_{ABCD} \, C_{AEFG} \, {\mathcal{T}}_{BCEHIJ} \, {\mathcal{T}}_{DHIFGJ}\right.  \nonumber \\
&&\left. +C_{ABCD} \, C_{AEFG} \, {\mathcal{T}}_{BCFHIJ} \, {\mathcal{T}}_{DEHGIJ} \,
+\, C_{ABCD} \, C_{AEFG} \, {\mathcal{T}}_{BCHEIJ} \, {\mathcal{T}}_{DFHGIJ} \right] \, + \, h.c. \nonumber \\
&&{\,}
\ea
where we have neglected the numerical coefficients of the terms and
left the metric tensors implicit. We are only interested in
operators quadratic in the gauge-field $A_a$. Immediately it follows
that the only contribution can arise via $[C^{(0)}]^2
[{\mathcal{T}}^{(1)}]^2$, so that all of the metric factors and the
Weyl factors are non-fluctuating. There are two types of
contributions: those with only $F_{ij}$ and those with $F_{ij}$ and
$\nabla_k F_{ij}$.

\subsubsection{$C^2 {\mathcal{T}}^2$ terms with no $\nabla F^+$}

The first term (see \cite{Paulos:2008tn}) is given by
\begin{equation}
C^2
{\mathcal{T}}^2=\frac{30240}{86016}G_{(0)}^{KA}G_{(0)}^{BL}G_{(0)}^{MC}G_{(0)}^{NF}
G_{(0)}^{PH}G_{(0)}^{GQ}G_{(0)}^{IR}G_{(0)}^{JS}\,C^{(0)}_{ABC}{}^D
\, C^{(0)}_{KLM}{}^E \, {\mathcal{T}}^{(1)}_{DGFHIJ} \,
{\mathcal{T}}^{(1)}_{ENPQRS} \,.
\end{equation}
The two-tensor
$G_{(0)}^{ka}G_{(0)}^{bl}G_{(0)}^{mc}C^{(0)}_{abc}{}^d \,
C^{(0)}_{klm}{}^e =C^{(0)}_{abc}{}^d \, C_{(0)}^{abce}$ is diagonal
and has no $S^5$ indices. The entries are of the form
$(u^5,u^5,u^5,u^5,u^6)$. Therefore, if we want to consider the
five-dimensional gauge-invariant operators coming from the above
equation, then it is clear that the $C^{(0)}_{abc}{}^d \,
C_{(0)}^{abce}$ term gives us simply $\hat{C}_{lmn}{}^i \,
\hat{C}^{lmnj}$, where $\hat{C}$ is the Weyl tensor evaluated on the
AdS-Schwarzschild space only. Now, the remaining piece is
\ba
&&
G_{(0)}^{NF}G_{(0)}^{PH}G_{(0)}^{GQ}G_{(0)}^{IR}G_{(0)}^{JS}{\mathcal{T}}^{(1)}_{DGFHIJ}
\, {\mathcal{T}}^{(1)}_{ENPQRS} \, . \nonumber
\ea
The factors of ${\mathcal{T}}^{(1)}$ can only give $F_{ij}$ and
$G_{kl}$, where everything now is in the AdS-Schwarzschild space.
Therefore the two operators that can be obtained after integrating
out the $S^5$ from this piece are simply
\begin{equation}
\hat{C}_{lmn}{}^i \, \hat{C}^{lmnj} \ g^{pr} \, F_{i r} F_{j p} \, ,
\end{equation}
and
\begin{equation}
\hat{C}_{lmn}{}^i \, \hat{C}^{lmnj}g_{i j}F^2 \, .
\end{equation}
Schematically, the contribution of such operators is $u^4 (u^2
F_{xz}^2+u^3 F_{x u}^2)$.

This manner of computation can then be extended to all of the
ten-dimensional operators of the form $C^2{\mathcal{T}}^2$. For
example, the next term in the list is given by is
\begin{equation}
C^2 {\mathcal{T}}^2=G_{(0)}^{GK}G_{(0)}^{HL}G_{(0)}^{IM}G_{(0)}^{JN}
\,C^{(0)}_{AB}{}^{CD} \, C_{(0)}^{ABEF} \, {\mathcal{T}}^{(1)}_{CDGHIJ}
\, {\mathcal{T}}^{(1)}_{EFKLMN} \,.
\end{equation}
To get the contribution of this operator to the five-dimensional
effective theory we must enumerate the various types of operators
coming from two factors of $F_{ij}$ and as many factors of the
metric as we need, as well as the factor $C^{(0)}_{st}{}^{ij} \,
C_{(0)}^{st kl}$, evaluated in AdS-Schwarzschild. These will have
the form
\begin{equation}
\hat{C}_{st}{}^{ij} \, \hat{C}^{st kl} \, F_{i j}F_{k l} \, , \quad
\, \hat{C}_{st}{}^{ij} \, \hat{C}^{st kl} \, g_{i k} \, g_{j l} \,
F^2 \, , \quad \hat{C}_{st}{}^{ij} \, \hat{C}^{st kl} g_{i k} \,
g^{m n} \, F_{j n} F_{l  m} \, ,
\end{equation}
where we have used the fact that $C_2^{i j k l}=C^{(0)}_{st}{}^{ij}
\, C_{(0)}^{st kl}$ obeys $C_2^{i j k l}=C_2^{k l i j}=-C_2^{j i k
l}=-C_2^{i j l k}$. The contribution of these operators is again of
the schematic form $u^4 (u^2 F_{xz}^2+u^3 F_{x u}^2)$.

The final type of term is that given by
\begin{equation}
C^2 {\mathcal{T}}^2=G_{(0)}^{LH}G_{(0)}^{IM}G_{(0)}^{JN} \,
C^{(0)}_{A}{}^{BCD} \, C_{(0)}^{AEFG} \,
{\mathcal{T}}^{(1)}_{BCEHIJ} \, {\mathcal{T}}^{(1)}_{DLMFGN} \,.
\end{equation}
We must now enumerate all the operators that will contain the
six-tensor $C^{(0)}_{a}{}^{bcd} \, C_{(0)}^{aefg}$ and two factors
of $F_{ij}$ and all the necessary metric factors. For example, we
have
\begin{equation}
\hat{C}_{a}{}^{bcd} \, \hat{C}^{aefg} \, g_{c f} \, g_{d g} \, F_{b
m} \, g^{m n} \, F_{n e} \, .
\end{equation}
Again, the explicit $u$-dependence of this operator is of the same
form as that of the previous two.

\subsubsection{$C^2 {\mathcal{T}}^2$ terms with $\nabla F^+$}

Let us extend this kind of analysis to the term in ${\mathcal{T}}^{(1)}$
which goes like $\nabla F^+$. In this case, we can now build
operators from $F_{ij}$ and $\nabla_k F_{ij}$. The analysis then
follows exactly as before. For example, from the Weyl tensor with
two up indices, we obtain
\begin{equation}
[C_{(0)}^2]^{A B} ({\mathcal{T}}^{(1)} {\mathcal{T}}^{(1)})_{ab}
\longrightarrow \hat{C}_{lmn}{}^i \, \hat{C}^{lmnj} g^{bf} \, g^{cg}
\, \nabla_i \, F_{bc} \, \nabla_j F_{fg}  \,.
\end{equation}
One can check that this operator gives us terms like $u^7
(\partial_z^2 A_x)^2$. We also get $u^8 (\partial_z \partial_u
A_x)^2$, and $u^9 (\partial_u^2 A_x)^2$. In principle, we may also
have terms that involve the Levi-Civita tensor, such as
\begin{equation}
\hat{C}_{lmn}{}^i \, \hat{C}^{lmnj} \, g_{ij} \, \epsilon_{abcde} \,
\nabla^a F^{bc} \, F^{de}  \,.
\end{equation}
The point here is that these operators enter with a very high power
dependence in $u$. The reason is that we need at least five factors
of $g^{a b}$ to contract the indices, which then means that this
operator will enter with at least $u^9$ in its coefficient, rendering it harmless.

Also, there are now terms with odd numbers of derivatives acting on
the gauge field. These come from the connection piece in the
covariant derivative $\nabla$. For example, from the above operator
we have
\begin{equation}
\hat{C}_{lmn}{}^i \, \hat{C}^{lmnj} \, g^{bf} \, g^{cg} \,
\Gamma^{s}_{ab} \, F_{sc} \, \nabla_j F_{fg}  \,.
\end{equation}
This term then gives us contributions of the form $u^7 \partial_z
A_x \, \partial_u  \partial_z A_x$. Of course, we may have other
contractions amongst the terms, but the overall effect is the same.
The crucial point is the high power of $u$ which enters into these
higher derivative operators. Another operator is given by
\begin{equation}
\hat{C}_{lmn}{}^i \, \hat{C}^{lmnj} \, g_{i j} \, g^{ab} \, g^{cf}
\, g^{dg} \, \nabla_a F_{cd} \, \nabla_b F_{fg}  \, .
\end{equation}
A simple counting of powers of $u$ reveals that these two
operators contribute at the same order.

Consider now the contribution of the $C_{(0)}^2$ term with four
indices up. One may be tempted to think that the $u$-dependence
drops, but that is not the case. Consider for example a situation
where the contractions are such that we have the following operator
\begin{equation}
\hat{C}_{st}{}^{ij} \, \hat{C}^{st kl} \, g^{a b} \, \nabla_a F_{ij}
\, \nabla_b F_{kl}  \, .
\end{equation}
By direct computation of $\hat{C}_{st}{}^{ij} \, \hat{C}^{st kl}$,
we may show that the contributions of this operator enter at the
same power in $u$. Again, operators with the Levi-Civita tensor are
not ruled out, so we may obtain
\begin{equation}
\hat{C}_{st}{}^{ij} \, \hat{C}^{st kl} \,  \epsilon_{abcij} \,
\nabla^a F^{bc} \, F_{kl}  \, .
\end{equation}
The index contractions imply that the least power with which this
operator contributes is then $u^9$, as before. The same argument
follows for the contribution of $C_{(0)}^2$ with six indices up.

\subsection{$C^3 \mathcal{T}$ terms}

The $C^3 {\mathcal{T}}$ term is uniquely given by
\begin{equation}
C^3 {\mathcal{T}}=C^{JKMN} \, {C}_{KL}{}^{RS} \, C_{J}{}^{PLQ} \,
{\mathcal{T}}_{MNPQRS} + h.c. \, .
\end{equation}
The compactification of this term will receive two types of
contributions. In the first, all the gauge-field dependence will
reside in ${\mathcal{T}}$. This term is then written as $C_{(0)}^3
{\mathcal{T}}^{(2)}$, where ${\mathcal{T}}^{(2)}$ here denotes the
part of ${\mathcal{T}}$ containing two powers of the gauge field.
The other type of contribution will be that where the quadratic
dependence on the gauge field is shared between the $C^3$ factor and
the ${\mathcal{T}}$ tensor, and we denote this by $C^{(1)} C_{(0)}^2
{\mathcal{T}}^{(1)}$. We begin with the former.

\subsubsection{$C_{(0)}^3 {\mathcal{T}}^{(2)}$}

This term is given by
\begin{equation}
C^3 {\mathcal{T}}=C^{JKMN} \, {C}_{KL}{}^{RS} \, C_{J}{}^{PLQ} \,
{\mathcal{T}}^{(2)}_{MNPQRS} \, .
\end{equation}
Again, the vanishing of the Weyl tensor on the $S^5$ means that the
compactification of this term is straightforward. The Weyl tensors
simply go to the AdS ones, so that all the indices in the above
expression become AdS indices. Moreover, ${\mathcal{T}}^{(2)}$ cannot
contain pieces of the form $\nabla F$, as is clear from the
definition of the tensor ${\mathcal{T}}$. Thus, the
${\mathcal{T}}^{(2)}$ piece is restricted to providing two factors of
$F_{ij}$, as well as factors of the metric tensor. A simple
five-dimensional operator resulting from the compactification would
then be
\begin{equation}
\hat{C}^{jkmn} \,  \hat{C}_{j}{}^{plq} \hat{C}_{kl}{}^{rs} \, g_{mp}
\,  g_{rq} \, g_{ns} \, F^2 \, .
\end{equation}
The resulting contribution is then $u^6 (u^2 F_{xz}^2+u^3
F_{xu}^2)$. Another operator is given by
\begin{equation}
\hat{C}^{jkm}{}_s \,  \hat{C}_{j}{}^{plq} \hat{C}_{kl}{}^{rs}
\,F_{mr } F_{pq}\, .
\end{equation}
We find this to be given by $u^8 (F_{xz}^2+u F_{xu}^2)$, exactly as
for the previous operator. There are many other contractions for the
indices inside this operator, but they all contribute at the same
order in $u$.

\subsubsection{$C_{(1)} C_{(0)}^2 T^{(1)}$}

First, write the following shorthand notation for the contraction of
three Weyl tensors
\begin{equation}
C^{JK}{}_{C}{}^{N} \,  C_{JE}{}^{LQ}
C_{KLH}{}^{S}=[C_{(0)}^3]_C{}^N{}_E{}^Q{}_H{}^S \, .
\end{equation}
In these terms the gauge-field dependence enters directly into the
Weyl tensor itself. These are in principle very complicated terms.
This can be schematically written as
\ba
&& G_{(0)}^{CM}G_{(0)}^{EP}G_{(0)}^{HR}\left[
[C_{(1)}C_{(0)}C_{(0)}]_C{}^N{}_E{}^Q{}_H{}^S +
[C_{(0)}C_{(1)}C_{(0)}]_C{}^N{}_E{}^Q{}_H{}^S+
[C_{(0)}C_{(0)}C_{(1)}]_C{}^N{}_E{}^Q{}_H{}^S \right] \nonumber \\
&& {\mathcal{T}}^{(1)}_{MNPQRS} \, .
\ea
where for example we have for the second term:
\begin{equation}
G_{(0)}^{CM}G_{(0)}^{EP}G_{(0)}^{HR}\left[  C_{(0)}C_{(1)}C_{(0)}
\right]_C{}^N{}_E{}^Q{}_H{}^S =G_{(0)}^{CM}G_{(0)}^{EP}G_{(0)}^{HR}
C_{(0)}^{JK}{}_{C}{}^{N} \,  C^{(1)}_{JE}{}^{LQ}
C^{(0)}_{KLH}{}^{S}\,.
\end{equation}
We raise the indices and consider the tensor $C_{(0)}^{JKM}{}_S \,
C^{(1)}_{J}{}^{PLQ} C^{(0)}_{KL}{}^{RS} $. The indices $J$, $M$,
$N$, $L$, $R$, $S$ are all AdS indices. To get a non-zero result, we
then require that both $P$ and $Q$ are internal or AdS indices. But
the tensor $C^{(1)}_{J}{}^{PLQ}$ is off-diagonal, and so $P$ and $Q$
cannot be AdS indices. Therefore, we can write the contribution of
this term as
\begin{equation}
C_{(0)}^{jkm}{}_s \,  C^{(1)}_{j}{}^{\tilde{p}l\tilde{q}}
C^{(0)}_{kl}{}^{rs} \, T^{(1)}_{mn\tilde{p}\tilde{q}rs} \, .
\end{equation}
where $\tilde{p}$ and $\tilde{q}$ are internal $S^5$ indices.
Examining the tensor $C^{(1)}_{j}{}^{\tilde{p}l\tilde{q}}$, we find
that it only has first derivatives of the gauge field $A_a$.  We
must now determine what manner of AdS-Schwarzschild tensors can come
from $C^{(1)}_{j}{}^{\tilde{p}l\tilde{q}}$ upon integrating out the
five-sphere. The fact that $C^{(1)}_{j}{}^{\tilde{p}l\tilde{q}}$
contains two internal indices means that the only available tensor
is again simply of the form $g^{la}F_{ja}$. For example, we cannot
obtain $\hat{R}^{la}F_{ja}$ because this one contains three
derivatives, but the maximum number of derivatives contained in the
Weyl tensor is two.

From the ${\mathcal{T}}^{(1)}_{mn\hat{p}\hat{q}rs}$ term we get the
usual suspects, namely terms like $F_{ij}$ and $G_{ab}$. In
principle, we may obtain terms of the form $\epsilon_{a b c d
e}\nabla^{a}F^{bc}$ as well. For example, we have the operator:
\begin{equation}
\hat{C}^{jkmn} \,  \hat{C}_{kl}{}^{rs}  \, g^{la} \, g_{nr} \,
F_{aj} \, F_{ms}\, .
\end{equation}
We may also have an operator like
\begin{equation}
\hat{C}^{jkmn} \,  \hat{C}_{kl}{}^{rs}  \, g^{la} \, g_{nr} F_{aj}
\, \epsilon_{msdef} \, \nabla^d F^{ef}\, .
\end{equation}
The latter operator actually contributes at a very high order in
$u$, because of the factors of the metric which contract the
Levi-Civita tensor with the $\nabla F$ tensor.

\subsection{The contribution of the operators as a series in $u$}

Having determined the form of the five-dimensional operators
descending from the eight-derivative corrections to type IIB supergravity,
we are able to compute their contribution to the gauge-field
Lagrangian explicitly. Given that the relevant quantity as far as
holography is concerned is the on-shell action evaluated on the
boundary of the space, it is sufficient to exhibit the low-$u$
dependence of the contributions. We find the following terms:
\ba\label{high-u-terms}
&& u^6 (\partial_\alpha A_\beta)^2 +\cdots +u^7 (\partial_u A_\beta)^2 \nonumber \\
&&+u^7 (\partial_\alpha A_\beta)(\partial_u \partial_\gamma A_\beta)
+ u^7(\partial_\alpha \partial_\gamma A_\beta)^2 \nonumber \\
&&+ u^8 ( \partial_u A_\beta)(\partial_u \partial_u A_\beta)+\cdots
+u^8 (\partial_\alpha \partial_u A_\beta)^2 +\cdots + u^9 (\partial_u^2 A_\beta)^2 +\cdots
\ea
where the $\cdots$ denote terms which have coefficients that contain
a higher $u$-dependence. The notation here is such that Greek
indices $\alpha, \beta, \gamma$ denote the four-dimensional
Minkowski slices of the AdS-Schwarzschild space, {\it i.e.} the
directions $t$, $x$, $y$, $z$. The crucial point is that the
inclusion of these terms does not affect the on-shell action of perturbations 
whose typical length scale is much smaller than $1/T$, as we discuss in the 
next section.

\section{The corrected equations of motion}
\label{finitelambda}

We have argued above that inserting the perturbed metric and
five-form field strength into the eight-derivative corrections
results in a slew of operators for the gauge field, all of which
contain at least two factors of the gauge field strength $F_{ab}$
and two factors of the AdS-Schwarzschild Weyl tensor
$\hat{C}_{a}{}^{bcd}$. In a certain sense the higher dimensionality
of the corrections is then replaced by high dependence on the radial
coordinate $u$. In principle, we should be able to organize the
result of the dimensional reduction of the
${\mathcal{O}}(\alpha'^3)$ corrections into a series of
eight-derivative gauge invariant operators quadratic in $F_{mn}$
with fixed coefficients, in the manner of
\cite{Ritz:2008kh,Myers:2009ij}. However, whereas the latter
references consider operators that are at most carrying six
derivatives, our case goes up to eight derivatives, and the large
number of such distinct operators, coupled with the sheer size of
the expression produced even just by the $C^4$ correction (see
Appendix A), means that such a programme is unfeasible. We therefore
rely on the results of the previous section, considering the terms
with the lowest $u$-dependence. Let us illustrate the behaviour of
these terms schematically: the action in the presence of the
corrections at small $u$ is given by
\ba
\textrm{action}& \propto &\int d^4x \, du \, \frac{1}{2u^3}
\left[-u^2 \left(\partial_t A_x\right)^2+u^2 \left(\partial_i A_x\right)^2
+\frac{4 u^3 r_0^2}{R^4}\left(\partial_u A_x\right)^2 \right. \nonumber \\
&& \left. + \gamma B_1 u^9  \left(\partial_u^2 A_x\right)^2
+  \gamma B_2 u^8 \left(\partial_u^2 A_x\right)
\left(\partial_u A_x\right) + \cdots \right] \, , \label{simple-corr}
\ea
where the first line is the contribution of the minimal $F^2$
kinetic term, and the terms with coefficients $B_i$ arise directly
from the eight-derivative term itself, with the dots denoting extra
terms arising from the eight-derivative corrections but containing
less $u$-derivatives (the terms considered here are the most
problematic). Due to the high positive power of $u$ in the terms
coming from $S_{C,{\mathcal{T}}}$, we find that none of the terms
produced affect the solutions of the equations of motion in the
ultraviolet. They all enter with at least $u^6$ in their
coefficient, rendering them irrelevant at small $u$. To see this,
recall that the relevant quantity for holography is the on-shell
action evaluated at the boundary of the space $u=0$. Now, it is easy
to show that the gauge field must behave like $A_a = a + b \, u + c
\, u \, \log(u) + \cdots$ near the boundary, because the ultraviolet
boundary is a regular singular point of the equations of motion,
with indices $\sigma=0,1$. If we take this form and plug it into
Eq.(\ref{simple-corr}), and then take the limit $u\to0$, the only
contributing terms will be those coming from the minimal kinetic
term. We refer the reader to Appendix B for more details of this
argument, and a thorough examination of the equations of motion. 
Further, this conclusion means that we may
eliminate all terms of ${\mathcal{O}}(u^4)$ or higher from $L(u)$,
$P(u)$ and $K(u)$ as defined in Eq.(\ref{proper-metric}), as they
will not contribute to the {\emph{on-shell}} action. The effect of
this is dramatic: it means that for the purposes of this computation
we can assume that the corrected metric is factorisable, because
$L(u)=1+{\mathcal{O}}(u^4)$, and therefore drops out of the entire
computation. 
A final important observation is that the overarching $SU(4)$ gauge symmetry of the vector
fluctuations ensures that, at least at quadratic order in the
Lagrangian, no mixing with other fluctuations can occur. For fields
which are sourced by the ${\cal {O}}(\alpha'^3)$ corrections themselves, any
effect on our calculations will contribute to even higher power of $\alpha'$ and
therefore does not enter into what follows \cite{Myers:2008yi}.

The upshot of the preceding arguments is that we may compute the
equations of motion solely using the minimal $F^2$ term and the
corrected metric, retaining only ${\mathcal{O}}(u^2)$ corrections in
the functions $P(u),\, K(u)$ and setting the warp factor $L(u)$ to
one, as it is given by $1+{\mathcal{O}}(u^4)+\cdots$. Given these
enormous simplifications, we may now present the
corrected equations of motion for the gauge fields. We first fix the gauge $A_u = 0$, and
choose the perturbation as a plane-wave propagating in the
$x_3$-direction \cite{Hatta:2007cs}. Thus, an appropriate {\it
ansatz} for the gauge field is
\be
A_{\mu}(t, \vec{x}, u) = e^{-i \omega t + i q x_3} \, A_\mu(u) \, .
\ee
The equations derived from Eq.(\ref{action}) are then given by
\ba
\varpi \, A_0' + \kappa \, f \, K^2(u) \, A_3' &=& 0 \, , \label{A3corrected} \\
&& \nonumber \\
A_i'' + \frac{f'}{f} \, A_i' + \,
\partial_u \left(\log{\left[ \frac{K(u) L^{7}(u)}{P(u)} \right]} \right) \, A_i' + \left[\frac{\varpi^2\,
- \kappa^2 \, f \, K^2(u)}{u \, f^2 K^2(u)}\right]P^2(u) A_i  \, &=& 0 \, , \label{Aicorrected} \\
&& \nonumber \\
A_0'' +  \,
\partial_u \left(\log{\left[\frac{L^{7}(u)}{P(u) \, K(u)}\right]}\right) \, A_0' -\frac{\kappa}{u \, f}
\, P^2(u) \, (\kappa \, A_0 + \varpi \, A_3) &=& 0 \, .
\label{A0corrected}
\ea
where we have defined $\varpi =\omega R^2/(2 r_0)$ and $\kappa =q
R^2/(2 r_0)$. Defining $\tilde{a}(u) \equiv A_0'(u)$ we may recast
Eq.(\ref{A0corrected}) into
\ba
\tilde{a}'' + \frac{(u \, f)'}{u \, f} \, \tilde{a}' + \partial_u
\left(\log{\frac{L^7(u)}{P^3(u)K(u)}}\right) \, \tilde{a}' \,+\,
\frac{P^2(u)}{u\, f^2} \, \left( \frac{\varpi^2- \kappa^2 \, f
K^2(u)}{K^2(u)} \right) \, \tilde{a}
&& \nonumber \\
&& \nonumber \\
+ \frac{P^2(u)}{uf} \, \partial_u \left(\frac{uf}{P^2(u)}
\partial_u \log{\left[\frac{L^7(u)}{P(u)K(u)}\right]}\right) \, \tilde{a}
&=& 0 \, .\label{acorrected}
\ea
In order to solve the above equations we have to impose certain
boundary conditions. For the non-vanishing $U(1)$ gauge fields we
have generic boundary conditions at $u=0$. Specifically, from
Eq.(\ref{A0corrected}) we obtain
\be
\lim_{u \rightarrow 0} \, [u \, \tilde{a}'(u)] = \kappa \, (\kappa
\, A_0 + \varpi \, A_3)|_{u=0} = \kappa^2 \, A_L(0) \, .
\ee
On the other hand, at $u=1$ the appropriate boundary condition that
must be imposed is equivalent to selecting only solutions that
describe waves going into the black hole, such that there is no
reflection off the horizon \cite{Son:2002sd, Policastro:2002se}. At
zero temperature, this condition is consistent with the requirement
of regularity of the solutions at the AdS horizon $u \rightarrow
\infty$.

Once we know the solutions of the equations of motion, the next
step is to evaluate the on-shell action by inserting the solutions
into Eq.(\ref{action}), which, after integration on $u$ and using
the boundary conditions above, gives \footnote{We remind the 
reader that, following the prescription of \cite{Son:2002sd,Herzog:2002pc,Hatta:2007cs}, 
we have dropped the contribution to the on-shell action coming from the horizon at $u=1$.}
\be
S_{on-shell} = - \frac{N^2 \, r_0^2}{16\pi^2 R^4} \, \int d^4x
\left[\tilde{a} \left(A_0 + \frac{\varpi}{\kappa} \, A_3
\right)|_{u=0} - A_i \,
\partial_u A_i(u)|_{u=0} \right] \, . \label{on-shell-action-4d}
\ee
Defining the on-shell action density
\be
S_{on-shell} = \int d^4x \, \hat{S}_{on-shell} \, ,
\label{5d-on-shell-action}
\ee
one may now obtain the desired current-current correlator by
differentiating with respect to the boundary value of the gauge
field $A_\mu \equiv A_\mu(u=0)$, so that
\be
R_{\mu\nu} = \frac{\partial^2 \hat{S}_{on-shell}}{\partial A_\mu \,
\partial A_\nu} \, .
\ee
The results of this section can therefore be used to compute the
fully-corrected solution of the equations of motion for the gauge
fields $A_\mu$ at order ${\mathcal{O}}(\alpha'^3)$, in any desired
regime of the parameters of the system, provided that the length scale 
of the perturbation is much smaller than $1/T$. By holography, this enables
us to obtain the behaviour of the electromagnetic current-correlator
at finite 't Hooft coupling. We now proceed to solve the equations of
motion of the gauge fields in the regime appropriate for deep inelastic scattering.

\subsection{Solving the bulk equations}

At this point we need to solve the Maxwell equations for the bulk
$U(1)$ gauge field. To avoid confusion, we define $\kappa_0=q/{2 \pi
T}$ and $\varpi_0=\omega/{2 \pi T}$. The quantities denoted with a
subscript $_0$ are those corresponding to the case where ${\cal
{O}}(\alpha'^3)$ corrections are not considered ({\it i.e.} infinite
't Hooft coupling limit). It is also convenient to define
${\mathcal{K}}^2 = \kappa^2 -\varpi^2$ (recall that the virtuality
$Q$ is given by $Q^2=q^2-\omega^2$, so ${\mathcal{K}}=Q R^2/(2
r_0))$. We also define
${\mathcal{K}}_0^2=\kappa_0^2-\varpi_0^2=Q^2/({2 \pi T})^2$. It is
then convenient to recast the EOMs as a time-independent
Schr\"odinger-like equation \footnote{We here neglect the terms in the Lagrangian coming from 
the higher-derivative corrections, as they do not influence the final result.}
\be
\psi'' - V(u) \, \psi = 0 \, . \label{schroedinger}
\ee
For this purpose we define the function $\psi(u) = \Omega(u) \,
\tilde{a}(u)$ and by choosing
\be
\Omega(u) = \left[\frac{u \, f(u) \, L^7(u)}{P^3(u) \,
K(u)}\right]^{1/2} \, ,
\ee
we obtain the Schr\"odinger-like equation (\ref{schroedinger}) with
the potential given by
\be
V(u)=\frac{\Omega''(u)}{\Omega(u)} - \left\{\frac{P^2(u)}{u f} \,
\partial_u \left(\frac{u f}{P^2(u)}\partial_u
\log{\left[\frac{L^7(u)}{P(u) \, K(u)}\right]}\right)+
\frac{P^2(u)}{u\, f^2} \, \left( \frac{\varpi^2- \kappa^2 \, f
K^2(u)}{K^2(u)} \right)\right\} .
\ee
Let us firstly try to intuitively understand the relation between the
gravity and field theory descriptions by analyzing the parametric
dependence of the potential barrier given by the above potential.

{{
\begin{center}
\epsfig{file=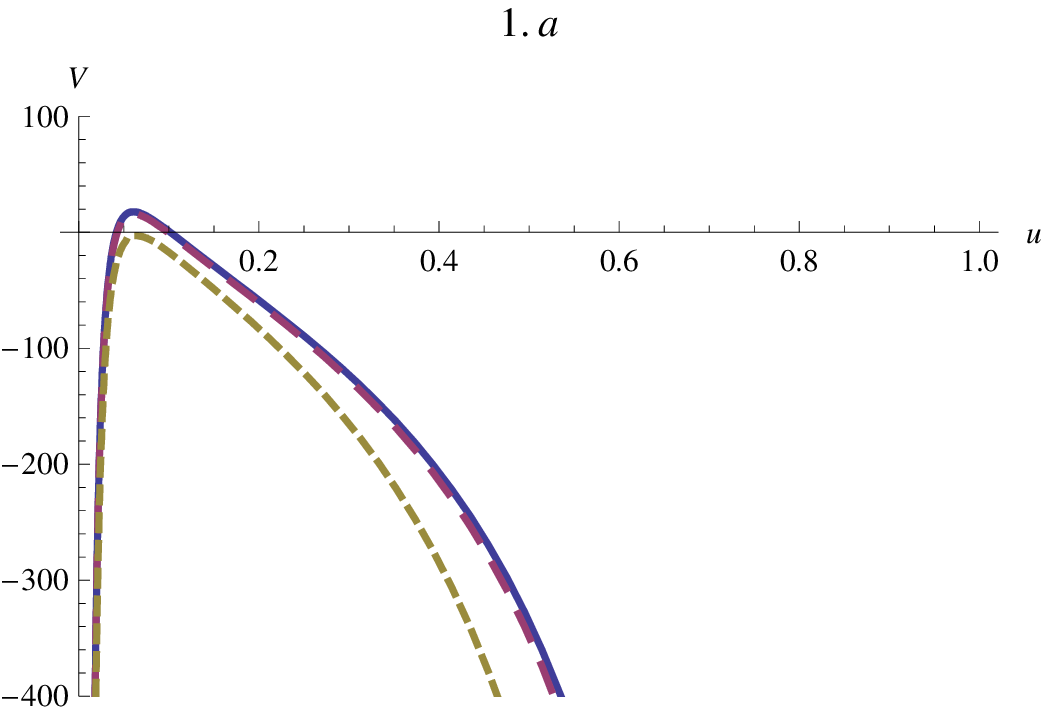, width=8.cm}
\end{center}

\begin{center}
\epsfig{file=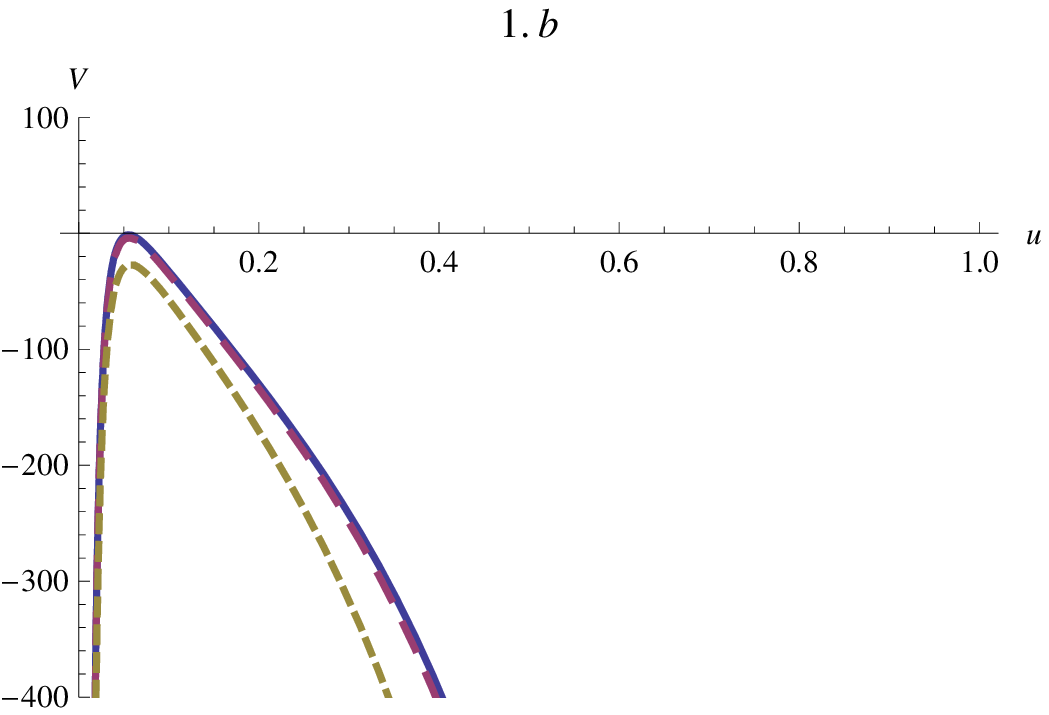, width=8.cm}
\end{center}


\begin{center}
\hspace{10.5cm} \epsfig{file=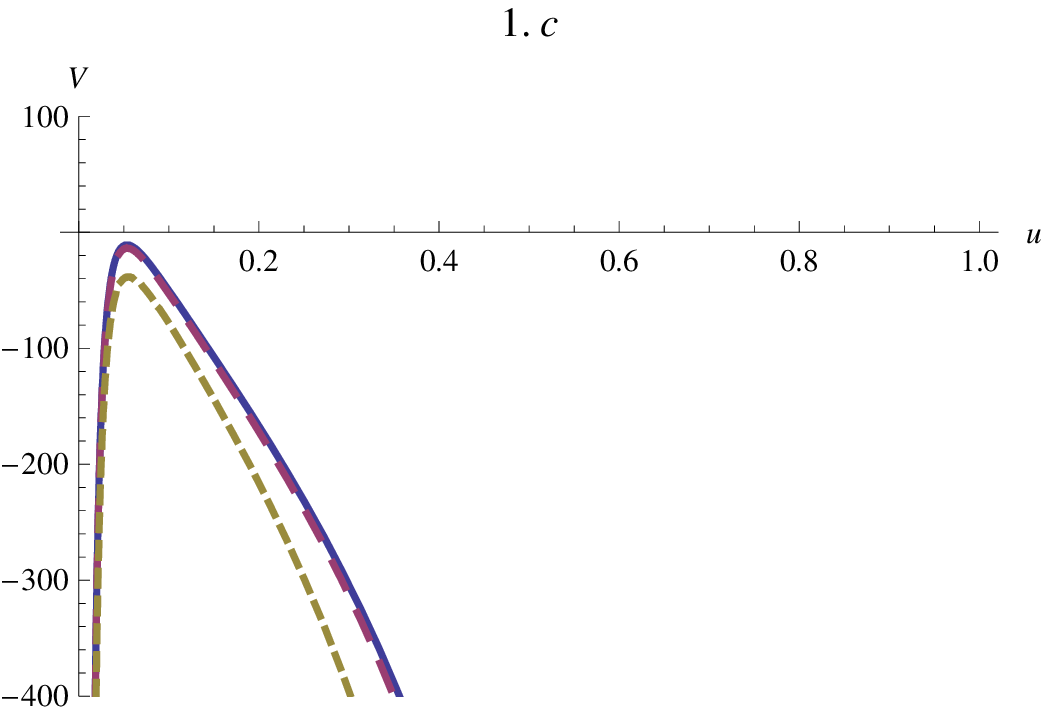, width=8.cm}
\end{center}

\centerline{{\small{Figure 1. The potential barrier for the $A_0$ gauge
field.}}} \vspace{1.cm} \baselineskip=20pt plus 1ptminus 1pt}


There are special regions of parameter space for which the physical
behaviour is rather distinct, and they are distinguished by the
following ratio, which we denote by $r_q$, and which is defined in
terms of the physical dimensionfull quantities of the plasma, namely
$\omega$, $q$ and $T$:
\be
r_q = \frac{\kappa_0}{{\mathcal{K}}_0^3}=\frac{q/{2 \pi T}}{(Q/{2 \pi
T})^{3}}= \frac{\kappa}{{\mathcal{K}}^3} \, \left( 1 +
\gamma\frac{265}{16}\right)^2 \, .
\ee
In Figure 1 we plot the ${\cal {O}}(\alpha'^3)$-corrected potential
barrier $V(u)$ as a function of the variable $u$, for different
parametric values of the ratio $r_q$ and different values of the 't
Hooft coupling, as explained below. We also plot the potential
barrier for the same values of this ratio without string theory
corrections, thus allowing us to see the effect of finite coupling
explicitly. In the limit of infinite coupling, the potential becomes
\be
\lim_{\lambda \rightarrow \infty} V(u) = \frac{1}{u(1-u^2)^2}
\left[-\frac{1}{4 u} \, (1 + 6 u^2 - 3 u^4) + {\mathcal{K}}_0^2
-\kappa_0^2 \, u^2 \right] \, .
\ee
The potential barrier for the longitudinal mode $A_0$ without
$\alpha'$ corrections is plotted with a solid line. We can
distinguish among three possible parametric situations in terms of
the ratio $r_q$. In the first case (figure 1.a) the ratio $r_q$ is
1.14, which gives a non-vanishing potential barrier. This case
corresponds to intermediate energies where the structure functions
of deep inelastic scattering are expected to be very small. For
extremely large values of $\lambda$ the "tunneling effect" through
the potential barrier is very small. For finite values of $\lambda$
the height of the potential barrier decreases, depending on the
actual value of $\lambda$, thus enhancing the tunneling effect as
the value of the 't Hooft coupling decreases. In fact, we have used
the values $\lambda=50$ (dashed line) and $\lambda=10$ (dotted line)
to show explicitly this effect in each figure. In the figure 1.b
$r_q=1.539$ and we see that for this limiting case the height of the
barrier vanishes for $\lambda \rightarrow \infty$. Figure 1.c shows
the potential for $r_q=1.71$. This case corresponds to the high
energy scattering process, where the potential barrier disappears
and the wave can propagate all the way towards the black hole
horizon and can thus be absorbed. This implies that the retarded
current-current correlation function acquires an imaginary part
thereby giving non-vanishing structure functions. We confirm that
the ${\cal {O}}(\alpha'^3)$-corrections to the potential decrease
the height of the barrier independently of the
$\kappa/{\mathcal{K}}^3$ values. This enhances the probability of
complete propagation of the wave to the black hole horizon,
therefore increasing the imaginary part of the tensor $R_{\mu\nu}$
as the value of $\lambda$ decreases. Notice that for the above
figures we have used the values $\lambda=50$ which gives $\gamma
\simeq 0.00042$, while for $\lambda=10$ it gives $\gamma \simeq
0.00474$. In addition, the height of the barrier is very sensitive
to the value of the 't Hooft coupling as can be seen from the
figures.

We now focus on the transvers modes $A_i(u)$. From
equation (\ref{Aicorrected}), we can define
\be
\phi(u) = \Sigma(u) \, A_i(u) \, ,
\ee
where
\be
\Sigma(u) =  \left(\frac{K(u)\,L^7(u)\,f}{P(u)}\right)^{1/2} \, .
\ee
Proceeding as in the previous case, we obtain a
time-independent Schr\"odinger-like equation for $A_i$
\be
\phi''(u) - V(u) \, \phi(u) = 0 \, ,
\ee
where the potential is
\be
V(u)=\frac{\Sigma''}{\Sigma}-\frac{P^2(u)}{u\, f^2} \, \left(
\frac{\varpi^2- \kappa^2 \, f K^2(u)}{K^2(u)} \right) \, ,
\ee
which for $\lambda\rightarrow\infty$ reduces to
\be
\lim_{\lambda\rightarrow\infty} V(u)=\frac{{\mathcal{K}}_0^2-u
\left(u \kappa_0^2+1\right)}{u \left(u^2-1\right)^2} \, .
\ee
{{\vspace{1.cm}

\begin{center}
\epsfig{file=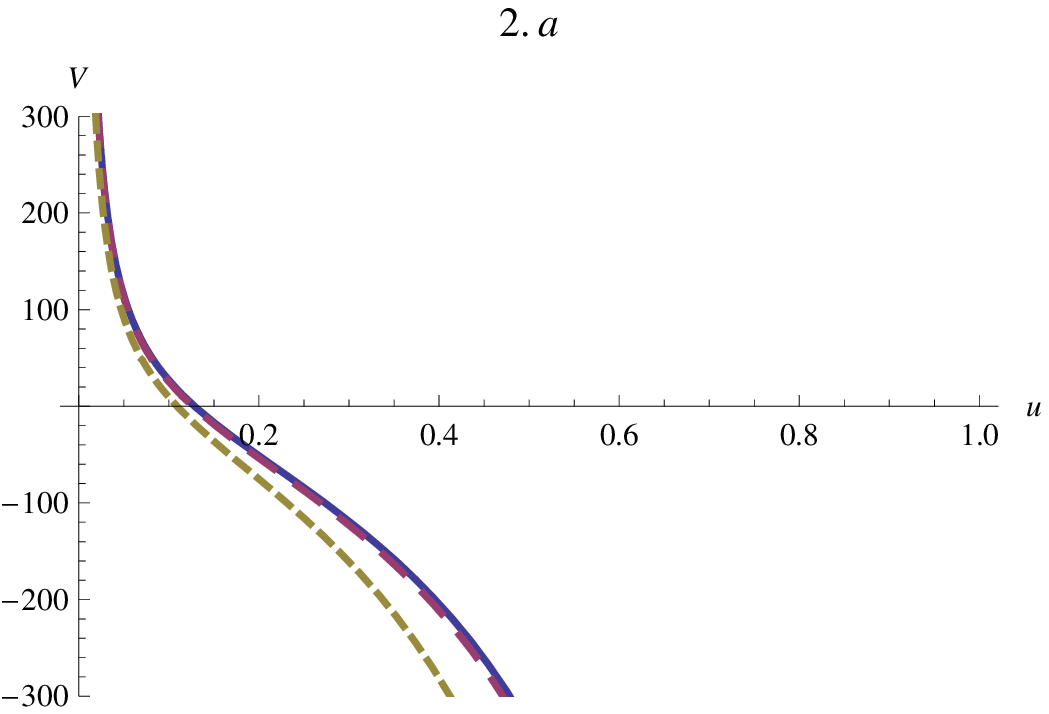, width=8.cm}
\end{center}

\begin{center}
\epsfig{file=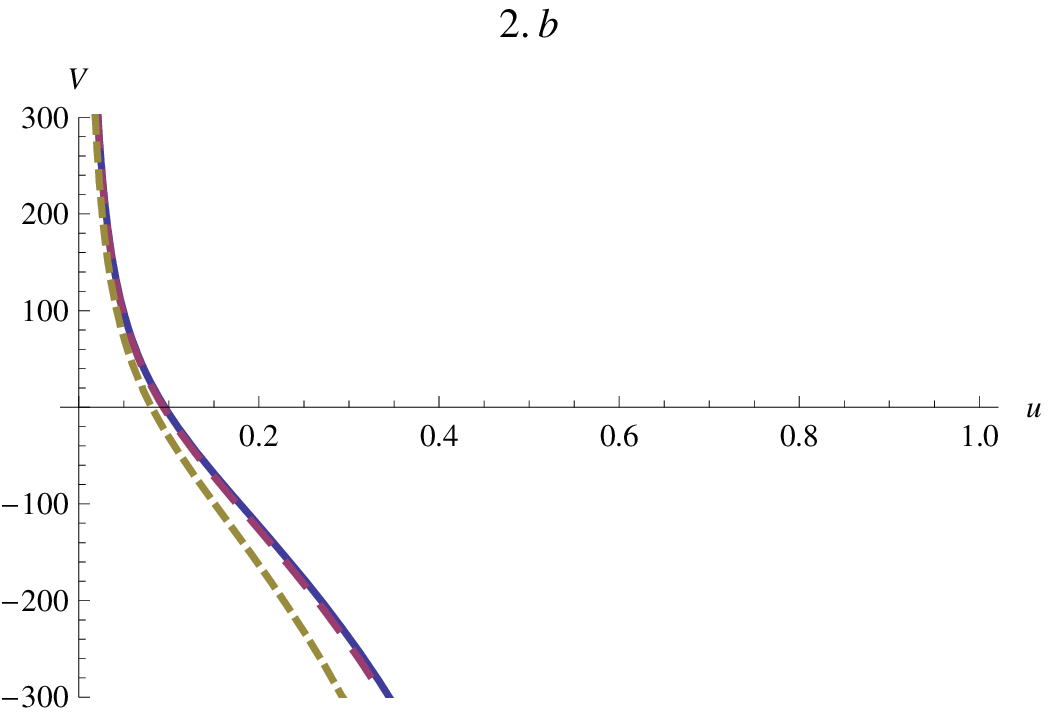, width=8.cm}
\end{center}

\begin{center}
\hspace{10.5cm} \epsfig{file=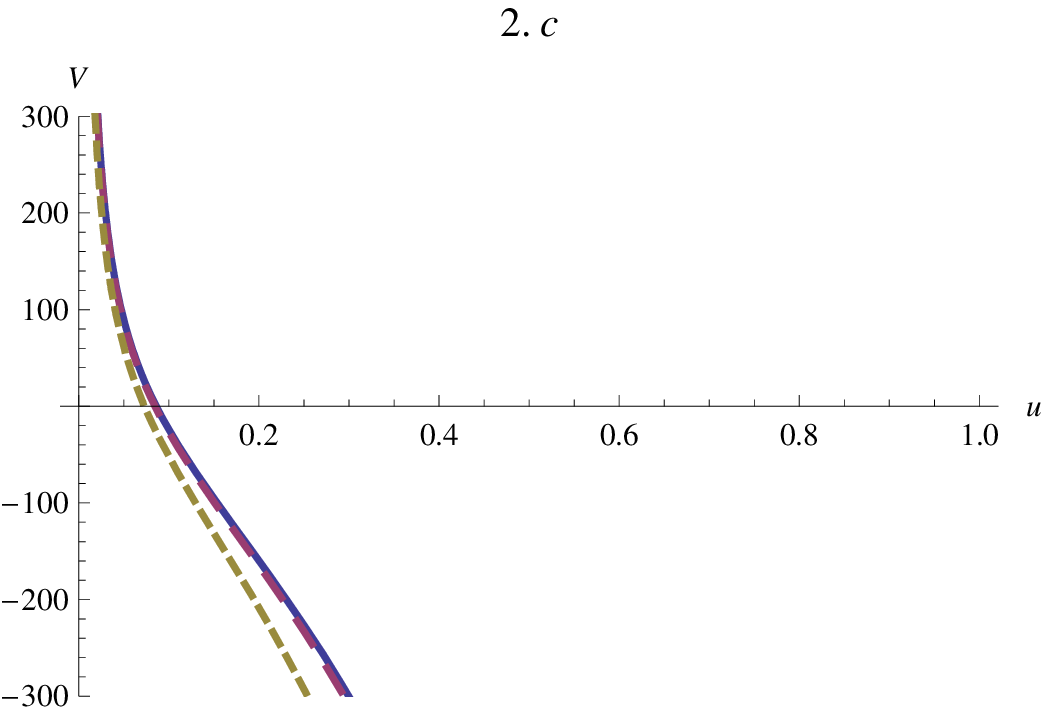, width=8.cm}
\end{center}

\centerline{{\small{Figure 2. The potential barrier for the transverse
modes.}}} \vspace{1.cm} \baselineskip=20pt plus 1ptminus 1pt}


In figure 2 we display the potential barrier for the transverse
modes $A_i$. From top to bottom, these figures correspond to the
ratios $r = 1.14, \, \,  1,539$ and  $1.71$, respectively. In each
figure we have three curves corresponding to $\lambda \rightarrow
\infty$ (solid line), $\lambda=50$ (dashed line) and $\lambda=10$
(dotted line) as in figure 1. It is clear that we again have a
reduction in the barrier height for decreasing $\lambda$, which
implies an enhancement of the structure functions.

Having physically motivated the expected enhancement in our results,
we now compute the current-current correlation functions, reading
off the structure functions from the imaginary parts. Interestingly,
one can consider two different parametric regimes.

Let us briefly study the low energy regime, leaving details of the
computation to the Appendix C. We focus on the parametric region
where $\kappa \ll {\mathcal{K}}^3$, which is equivalent to the low
temperature regime $q T^2 \ll Q^3$. In addition, by restricting the
radial coordinate $u$ to a small region $0 \leq u \leq
1/{\mathcal{K}}^2 \ll 1$, equations (\ref{Aicorrected}) and
(\ref{acorrected}) can be solved in perturbation theory. The
physical interpretation is given by a multiple scattering series at
low energy \cite{Hatta:2007cs}.

The on-shell 5D action of Eq.(\ref{5d-on-shell-action}) together
with the on-shell action density (\ref{on-shell-action-4d}) can be
split into two terms:
\be
S_{on-shell} = S^{(0)}_{on-shell} + S^{(1)}_{on-shell} \, ,
\ee
where $S^{(0)}_{on-shell}$ is the zero temperature contribution to
the on-shell action, while the other term is proportional to $T^4$.
Thus, within the parametric region mentioned above and with the
regularization scheme used in \cite{Hatta:2007cs} the on-shell
action becomes
\be
S^{(0)}_{on-shell} = - \frac{N^2}{64 \pi^2} \,
\log{\left(\frac{Q^2}{\Lambda^2} \right)} \, \left[ \left(q \, A_0 +
\omega \, A_3 \right)^2 - Q^2 \, {\mathcal{A}}_T \cdot
{\mathcal{A}}_T \right]_{u=0}  \, ,
\ee
where $\Lambda$ is a regulator in the gauge theory. As expected,
this expression is not corrected by the effect of the ${\cal
{O}}(\alpha'^3)$ term in the 10D action. Therefore, the
corresponding expression for the retarded current-current correlator
at zero temperature gets no $\alpha'$-corrections. For
$R_{\mu\nu}^{(0)}$ we obtain
\be
R_{\mu\nu}^{(0)}  = \frac{N^2 Q^2}{32 \pi^2} \,
\log{\left(\frac{Q^2}{\Lambda^2} \right)} \, \left(\eta_{\mu\nu} -
\frac{q_\mu q_\nu}{Q^2} \right) \, ,
\ee
which is real. These are indeed the expected results since at zero
temperature the AdS$_5 \times S^5$ metric is left uncorrected by the
higher derivative corrections to the classical supergravity action \cite{Banks:1998nr}.

The second term in the on-shell action above is
\be
S^{(1)}_{on-shell} \simeq \frac{N^2 \pi^2 T^4}{30} \,
\frac{q^2}{Q^6} \, (1 + 15 \gamma) \, \left[ \left(q \, A_0 + \omega
\, A_3 \right)^2 + \frac{3}{2} Q^2 \, {\mathcal{A}}_T \cdot
{\mathcal{A}}_T \right]_{u=0}  \, .
\ee
We thus see that $R_{\mu\nu}^{(1)}$ is corrected at finite 't Hooft coupling. 
The $\alpha'$-corrections do not introduce an imaginary part into the retarded current-current
correlator for low energies and, therefore, the plasma structure
functions in the present regime vanish\footnote{There is, however, a
very small contribution to the structure functions in this regime
which is due to the ``barrier tunneling" effect, which is similar to
the contribution reported in \cite{Hatta:2007cs}.}.

In the next section we explore the role of the $\alpha'$-corrections in the high
energy regime, where deep inelastic scattering is expected to occur.

\subsection{High-energy scattering}

In this section we consider the high energy regime where $\kappa \gg {\mathcal{K}}^3$.
We examine the equations of motion of the gauge fields, and we keep
only the leading terms in an expansion in powers of $u$. 
The idea is to compute the corrections to the equations of motion in a double expansion in $u$
and $\gamma$, simultaneously applying the condition $\kappa \gg {\mathcal{K}}^3$.
This is easy to do in practice, and a quick calculation reveals that the corrected
equations of motion are given by
\be\label{a-equation}
a(u)'' +\left[\frac{1}{u}+ {\mathcal{O}}(u)\right]a'(u)+
\left(\left[1+\frac{325}{4}\gamma\right] \kappa^2 u
+ {\mathcal{O}}(u^2)\right) a(u) =0 \,\, ,
\ee
and
\be\label{A-equation}
{\partial_u}^2 A_i(u)+
\left(\left[1+\frac{325}{4}\gamma\right] \kappa^2 u
+ {\mathcal{O}}(u^2)\right) A_i(u) =0 \, .
\ee
The analysis is then simplified considerably if one defines the
variable $k=(1+ \gamma \, 325/8)\kappa$, where it is crucial to keep
in mind the relations $\kappa =q R^2/(2 r_0)$ and $r_0 = \pi T
R^2/(1+ \gamma \, 265/16)$. 

Let us first consider the solution of Eq.(\ref{a-equation}). Defining $\xi = (2/3) k u^{3/2}$, 
we find that the general solution is given by $a(\xi)=c_1 J_0(\xi)+c_2 N_0(\xi)$. 
To fix the constants $c_{1,2}$, we must impose the appropriate boundary conditions. 
At the U.V. we have the generic Dirichlet condition demanded by AdS/CFT. 
At the I.R. we must impose the incoming wave boundary condition. One notices that an important simplification
occurs \cite{Hatta:2007cs}: it is the fact that this infrared b.c. can be imposed at relatively
small values of $u\ll1$. The argument is that, at high energies, the absence
of a potential barrier means that there is no mechanism to generate reflected
waves at intermediate values of $u<1$, since it would necessarily
describe reflection off the black hole. We proceed as follows: notice firstly 
that although $u\ll1$ the argument of the Bessel functions is large since $\xi = (2/3) k
u^{3/2}$ and $k\gg K^3$, for all the values far beyond the peak of the
potential $u\gg u_0 \approx 1/k^{2/3}$. In that region one can
asymptotically expand the Bessel functions and get: $J_0(\xi)
\approx \sqrt{2/(\pi \xi)} \cos(\xi-\pi/4)$ and $N_0(\xi) \approx
\sqrt{2/(\pi \xi)} \sin(\xi-\pi/4)$. Remembering the time-dependence $e^{-i \omega t}$, 
we see that the solution becomes an outgoing wave if $c_1=-ic_2$. We have thus fixed the ratio of 
$c_1$ to $c_2$, and the magnitude of the latter is then fixed by the U.V. boundary condition.
The same arguments can be applied to solve the equation of $A_i$, as explained in \cite{Hatta:2007cs}. 

We finally have that the solutions for equations (\ref{a-equation}) and (\ref{A-equation}) which obey the incident-wave condition at the black hole horizon
$u=1$ and the conditions demanded by the AdS/CFT correspondence at the boundary $u=0$ are given by \cite{Hatta:2007cs}
\be
a(u)=-i\frac{ \pi}{3} \, k^2 \,H_{0}^{(1)}\left(\frac{2}{3} k
u^{3/2}\right) {\mathcal{A}}_L(0) \, ,
\ee
and
\be
A_i(u)=\frac{i \pi}{\Gamma(1/3)}\left(\frac{k}{3}\right)^{1/3} \,
\sqrt{u} \, H_{1/3}^{(1)}\left(\frac{2}{3} k u^{3/2}\right)
{\mathcal{A}}_T(0) \, ,
\ee
where $H_{\nu}^{(1)}(x)$ is the first Hankel function defined by
$H_{\nu}^{(1)}(x)=J_{\nu}(x)+i Y_{\nu}(x)$, where $J_{\nu},Y_{\nu}$
are the usual Bessel functions of order $\nu$. The next step is to
evaluate the on-shell action for these field configurations. Using
the following form for the action density
\be
 \hat{S}_{on-shell} \,= - \frac{N^2 \, r_0^2}{16\pi^2 R^4} \,\left[\tilde{a}
\left(A_0 + \frac{\varpi}{\kappa} \, A_3 \right)|_{u=0} - A_i \,
\partial_u A_i(u)|_{u=0} \right] \, ,
\ee
we obtain
\ba
 \hat{S}_{on-shell} \, &=& - \frac{N^2 \, r_0^2}{48\pi^2 R^4}
 \nonumber \\
&& \, \left\{\,
k^2\left[2\left(\xi_E+\ln\left(\frac{k}{3}\right)\right)
 - i\pi\right]{\mathcal{A}}_L^2(0)+\frac{9\pi}{\Gamma^2(1/3)}
 \left(\frac{k}{3}\right)^{2/3}\left[\frac{1}{\sqrt{3}}-i \right]
 {\mathcal{A}}^2_T(0) \right\} \, , \nonumber \\
&&
\ea
where $\xi_E$ is the Euler-Gamma function $\xi_E=0.57\cdots$.
Remembering that
\be
F_1= \frac{1}{2 \pi} \, \textrm{Im} R_1 \, ,
\ee
we can express the structure function $F_1$ in terms of the
physical momentum $q$ and the temperature $T$, obtaining
\be
F_1 \simeq \left(1+\frac{5}{8}\xi(3) \lambda^{-3/2}\right)\, \frac{3 N^2
T^2}{16 \Gamma^2(1/3)} \left(\frac{q}{6 \pi T}\right)^{2/3}  \, ,
\ee
where $\lambda$ is the 't Hooft couppling and $\xi(3)\approx 1.20$.
Observe that our result is enhanced in comparison with the
zero-order result of \cite{Hatta:2007cs}. From this we can define
the transverse structure function as $F_T \equiv 2 x_B F_1$, where
$x_B= Q^2/(2 \omega T)$.

Now, we can similarly obtain the longitudinal structure function
$F_L$
\be
F_L \simeq \left(1+\frac{325}{32} \xi(3) \lambda^{-3/2}\right) \,
\frac{N^2 Q^2 x_B}{96 \pi^2}
 \, ,
\ee
which is related to  $F_2$ through $F_L \equiv F_2 - 2 x_B F_1$,
where
\be
F_2= \frac{(- n \cdot q)}{2 \pi T} \, \textrm{Im} R_2 =
\frac{\omega}{2 \pi T} \, \textrm{Im} R_2\, .
\ee
The parametric estimates of the above equations are similar to those
of \cite{Hatta:2007cs} where they do not include finite 't Hooft
coupling corrections. Thus, in our case we have
\be
F_T \propto \left(1+\frac{5}{8}\xi(3) \lambda^{-3/2}\right) \,
\frac{N^2 T^2}{x_B} \, \left( \frac{x_B^2 Q^2}{T^2} \right)^{2/3}
 \, ,
\ee
\be
F_L \propto \left(1+\frac{325}{32} \xi(3) \lambda^{-3/2}\right) \,
\frac{N^2 T^2}{x_B} \, \left( \frac{x_B^2 Q^2}{T^2} \right)
 \, .
\ee
We see that in the small-$x_B$ regime at $x_B\ll T/Q$, $F_L\ll F_T$.  As
in \cite{Hatta:2007cs}, this result looks quite different from the
results of DIS from a single hadron at weak and strong coupling,
where in the high energy limit the transverse and longitudinal
structure functions are parametrically of the same order.

\section{Conclusions}

In this work we have investigated the behaviour of holographic vector 
current-current correlators when the leading higher derivative corrections 
to the low energy type IIB supergravity are included. These corrections enter 
at order $\gamma \propto \alpha'^3$, and are built out of both the metric and the five-form
field strength. The metric at leading order in $\gamma$ is not
corrected by the higher-derivative terms containing $F_5$, but only
by the $\gamma \, C^4$ term.

By considering vector perturbations of the metric and the five-form
field strength around the corrected AdS-Schwarzschild background, we
derive the modified Maxwell equations for the Abelian gauge field
which is dual to the vector current of a gauged $U(1)$ subgroup of the 
${\mathcal{R}}$-symmetry group of ${\mathcal{N}}=4$ SYM at finite 
temperature\footnote{For the zero temperature case one recovers the results of
the low energy type IIB supergravity, {\emph{i.e.}} when no
$\alpha'$ corrections are included. This is in full agreement with
expectations, since at zero temperature the system reduces to the
simpler AdS$_5 \times S^5$ metric solution.}. Although the higher
curvature corrections induce a large number of additional operators
to the gauge field Lagrangian ({\emph{i.e.}} additional to the
minimal kinetic term), a careful examination reveals that these
operators enter with very high powers of the radial coordinate $u$
($u^6$ or higher). Given that the on-shell action for the gauge
field reduces to a boundary term to be evaluated in the ultraviolet
$u \to 0$, we thus find that the {\emph{direct}} effect of the extra
operators vanishes on-shell. The net result is that the influence of the higher-curvature string theory corrections is indirect, and is
effected by the modification to the metric only. We would like to
emphasize this result: the higher derivative operators induced in
five dimensions {\emph{ do not}} affect the on-shell action in the high energy (scattering) regime. The
latter is only affected by the modifications of the metric through
the minimal gauge-field kinetic term.

It is interesting to emphasize the distinct effects of the two sets
of ${\cal {O}}(\alpha'^3)$ corrections. On the one hand, there is
the $\gamma W_4=\gamma C^4$ term which only involves the Weyl tensor
and thus the metric. This term couples to the Einstein equations so
that its presence modifies the metric, yielding the corrections
computed in \cite{Gubser:1998nz,Pawelczyk:1998pb}. The
supersymmetric completion of the $C^4$ term brings a full set of
${\cal {O}}(\alpha'^3)$ corrections which contain the Ramond-Ramond
five-form field strength, as discussed above. These do not modify
the background metric, as discussed in
\cite{Paulos:2008tn,Myers:2008yi}. At the quadratic level for the
vector fluctuations, we have argued that the full set of
ten-dimensional higher-curvature terms only produce an indirect
effect on the on-shell action for the gauge fields. 
One may speculate that this simplification is a consequence of 
the maximal symmetry of the problem at hand.
Clearly, the case of
${\mathcal{N}}=4$ SYM is very special, in that the $S^5$
has vanishing Weyl tensor, so that all of the higher-derivative operators we found
contained at least two powers of the AdS-Schwarzschild Weyl tensor.
Moreover, the tensor ${\mathcal{T}}$ vanishes for the
AdS-Schwarzschild $\times S^5$ background, which also simplified our
analysis considerably and eliminated a large class of
five-dimensional operators. This is not necessarily the case for
other internal manifolds, implying that for complicated cases one
may actually find five-dimensional operators that directly influence
the holographic correlation functions. 

We have applied our results to compute the structure functions
governing deep inelastic scattering off an ${\cal {N}}=4$ SYM
plasma. We have found an enhancement of all of the relevant
structure functions. The same trend is found in deep inelastic
scattering off a single hadron \cite{Polchinski:2002jw}.

There are other interesting applications that can be addressed with the results obtained here. Among
them would be a computation of the leading 't Hooft coupling
corrections to the electric conductivity and the charge diffusion
constant of strongly-coupled ${\mathcal{N}}=4$ SYM. We will report
on these issues in a forthcoming paper \cite{us-hydrodynamics}.

~

~

~

\centerline{\large{\bf Acknowledgments}}

~

We thank Joe Conlon, Yang-Hui He, Adri\'an Lugo, Andre Lukas, Juan
Maldacena, John March-Russell, Carlos N\'u\~nez and Andrei Starinets
for useful discussions. We are very grateful to Carlos N\'u\~nez for a critical
reading of the manuscript. B.H. thanks Christ Church College, Oxford,
for financial support. The work of M.S. has been partially supported
by the CONICET and the ANPCyT-FONCyT Grant PICT-2007-00849.

\newpage

\appendix
\section{Appendix: Vector fluctuations and the $W_4$-term}

The fluctuation {\it ansatz} of Eq.(\ref{metric-ansatz}) and that of
Eq.(\ref{F5ansatz}) ensure that we pick a specific $U(1)$ subgroup
of the $SU(4)$ ${\mathcal{R}}$-symmetry group of the dual field
theory. Plugging the {\it ans${\ddot{a}}$tze} into the two-derivative Lagrangian
of type IIB supergravity gives the gauge kinetic term for the $U(1)$
field, as explained in section \ref{finitelambda}. The substitution
of the {\it ansatz} into the eight-derivative operators of
Eq.(\ref{10DWeyl}) results in a large expression, which, in
principle, can be placed into a series of higher-derivative
operators quadratic in the $U(1)$ field strength $F_{mn}$ (we have
ignored operators with higher factors of the field strength, as they
do not contribute to the linearized equations of motion), as we
discussed in section \ref{finitelambda}. Unfortunately, such a
scheme is made prohibitively difficult by the sheer size of the
expression. We here include the expression obtained when the  metric {\it ansatz}
is inserted into the $C^4$ term, to confirm the arguments of section
\ref{finitelambda}. Setting $A_u=0$ using the gauge symmetry, we
first decompose the $W_4$ term as follows,
\be
W_4 \, = \, \frac{u^6}{6 \pi ^6 f(u)^2 R^8 T^6} \left[
W_{xx}+W_{yy}+W_{tt}+W_{zz}+W_{zt} \right] \, ,
\ee
where $W_{\alpha\beta}$ contains only the quadratic combination
$A_\alpha A_\beta$. Observe that, at least for the $C^4$ term
computed here, the $A_x$ and $A_y$ perturbations do not mix with any
others, so that symmetry entails that $W_{yy}$ can be obtained from
$W_{xx}$ simply by replacing $A_x \to A_y$ in the latter. In what
follows we denote the various derivatives acting on gauge fields
using the notation $A_\mu^{(l,m,n)}(u)=\partial_t^l \partial_z^m
\partial_u^n A_\mu(t,z,u)$. In this appendix we exhibit the results
for $W_{xx}$ and $W_{tt}$ by way of illustrating the type of operators that are
induced. For $W_{xx}$ we obtain:
\ba
W_{xx}&=& \Bigg(64 \pi ^4 T^4 uf(u)^2 \left(78 u^4-67 u^2+14\right)  A_x^{(0,0,1)}(u)^2 +528 \pi ^4 T^4 u^{3}f(u)^4A_x^{(0,0,2)}(u)^2   \Bigg. \nonumber \\
&&+16 \pi^2 T^2 f(u)^2\left(7u^2-12\right)A_x^{(0,1,0)}(u)^2 - 704  \pi^2 T^2 uf(u)^3 A_x^{(0,1,0)}(u)A_x^{(0,1,1)}(u)\nonumber \\
&&- 384 \pi^2 T^2 u^2f(u)^3 A_x^{(0,1,1)}(u)^2-120 \pi^2 T^2 u^2f(u)^3 A_x^{(0,0,2)}(u)A_x^{(0,2,0)}(u) \nonumber\\
&&+ 16 \pi^2 T^2 \left(21u^4-45 u^2+16\right) A_x^{(1,0,0)}(u)^2+ 33  u f(u)^2 A_x^{(0,2,0)}(u)^2 \nonumber\\
&&+96  u f(u) A_x^{(1,1,0)}(u)^2- 128 \pi^2 T^2 u^2 f(u)^2 A_x^{(1,0,1)}(u)^2 \nonumber\\
&&- 64 \pi^2 T^2 uf(u)\left(u^2+3\right) A_x^{(1,0,0)}(u)A_x^{(1,0,1)}(u)+33 u A_x^{(2,0,0)}(u)^2\nonumber \\
&&+ 2 uf(u)\left[68 \pi^2T^2uf(u)A_x^{(0,0,2)}(u)+15 A_x^{(0,2,0)}(u)\right] A_x^{(2,0,0)}(u)    \nonumber \\
&&+16 \pi ^2 T^2 uf(u) \bigg\{ f(u) \left[ 4\pi^2T^2 uf(u) \left(28-53u^2\right)A_x^{(0,0,2)}(u)+5 \left(7u^2-4 \right) A_x^{(0,2,0)}(u) \right] \bigg. \nonumber \\
&&\bigg.\Bigg.- \left(12- 37 u^2 \right)A_x^{(2,0,0)}(u)
\bigg\}A_x^{(0,0,1)}(u) \Bigg) \, .
\ea
For the contribution quadratic in $A_t$, we get
\ba
W_{tt}&=& 384 \pi ^2 T^2 \, u^2 \,f^2(u)\, A_t^{(0,1,1)}(u)^2 \, + \, 7 u f(u)  A_t^{(0,2,0)}(u)^2\, +\, 33 u A_t^{(1,1,0)}(u)^2
\nonumber \\
&& + 228 \pi ^2 T^2 u^2 f(u)
A_t^{(1,0,1)}(u)^2  - 912 \pi^4 T^4
 u^3 f(u)^3 A_t^{(0,0,2)}(u)^2  \nonumber \\
&& +960 \pi ^4 T^4 u f(u)^2
\left(7 u^2-2\right) A_t^{(0,0,1)}(u)^2 + 184 \pi^2 T^2 u^2 f^2(u) A_t^{(0,2,0)}(u) A_t^{(0,0,2)}(u)\nonumber \\
&&+64 \pi^2 T^2 u \left[f(u)(5-7u^2) A_t^{(0,1,0)}(u) A_t^{(0,1,1)}(u) \right. \nonumber \\
&&\left. + 5 \left( -12 \pi^2 T^2 u f(u) A_t^{(0,0,2)}(u)+A_t^{(0,2,0)}(u)\right)A_t^{(0,0,1)}(u) \right] \, .
\ea
The other contributions are similar in structure, so we shall not present them here.

\section{Appendix: The equations of motion}

As we saw in the last section, the eight-derivative
${\mathcal{O}}(\alpha'^3)$ corrections introduce a multitude of
higher derivative operators, and we must take account of them
properly to solve the equation of motion within perturbation theory.
The situation is entirely analogous to that studied by Buchel, Liu
and Starinets in \cite{Buchel:2004di}. In that paper, the authors
were concerned with the tensor perturbations of the metric, but the
logic is the same. When we derive the equations of motion from an
action which contains higher derivative terms like $A_x'' A_x''$,
dangerous boundary terms like $\delta A_x'$ and $\delta A_x''$ will
be introduced. These threaten to ruin the consistency of the
variational principle, necessitating the addition of boundary
localized terms that ensure that all variations are simply
proportional to $\delta A_x$, so that the Dirichlet problem is
well-posed. Such an idea is familiar from Einstein gravity, where
the problem is made consistent by adding the exterior curvature
term. In our case, as in \cite{Buchel:2004di}, we must add
{\emph{boundary}} terms by hand to make the variational procedure
consistent. At the end of the day, these terms do not contribute to
the physical answers we seek in this work, as will become clear
shortly.

To see how this works in detail, it is convenient to introduce the
Fourier transform of the field $A_x$
\be
A_{x}(t, \vec{x}, u) = \int \frac{d^4k}{(2 \pi)^4} \, e^{-i \omega t + i q z} \, A_k(u) \, .
\ee
Upon inserting this into the total action for the gauge field $A_x$
({\it i.e.} the action containing both the two derivative term $F^2$
and the higher-curvature terms in the appendix above), we obtain the
expression:
\ba
S_{\textrm{total}} &=&- \frac{N^2 r_0^2}{16 \pi^2 R^4}
\int \frac{d^4k}{(2 \pi)^4} \int_0^1 du \,
\left[\gamma A_W A_k'' A_{-k}+ (B_1+\gamma B_W) A_k' A_{-k}' \right. \nonumber \\
&&\left. + \gamma C_W A_k' A_{-k}+(D_1+\gamma D_W) A_k A_{-k}
+\gamma E_W A_k'' A_{-k}''+\gamma F_W A_k'' A_{-k}' \right]  \, . \label{AxAction}
\ea
The coefficients $B_1$ and $D_1$ arise directly from the minimal
kinetic term $F^2$. The subscript $W$ indicates that the particular
coefficient comes directly from the eight-derivative corrections.
Moreover, $B_1$ and $D_1$ contain some $\gamma$-dependence, but they
are non-vanishing in the $\gamma \to 0$ limit, while every other
coefficient vanishes in that limit. We will discuss the explicit
form of the various coefficients shortly. First, we remark that upon
varying the action of Eq.(\ref{AxAction}), we obtain the equations
of motion:
\be
A\, A_k''+C A_k'+2 D A_k-\partial_u\left(2 B A_k'+C A_k +F A_k'' \right)
+\partial_u^2 \left(A A_k+2 E A_k''+F A_k'\right)\, = 0 \, ,\label{FullEOM}
\ee
where $B=B_1+\gamma B_W$ and so on. We may write this action in the form
\be
A_k''+p_1 A_k'+ p_0 A_k \, = {\mathcal{O}}(\gamma) \, ,
\ee
where we have moved all $\gamma$-dependence to the right. A careful
examination of the variation of Eq.(\ref{AxAction}) then convinces
us that in order to remove all dangerous variations to order
$\gamma^2$ we must add the boundary term
\be
S_{b} =- \frac{N^2 r_0^2}{16 \pi^2 R^4} \int \frac{d^4k}{(2 \pi)^4}
\int_0^1 du \, \partial_u\left[-\gamma A_W \, A_k A_{-k}'+\gamma E_W
\left(p_1 A_k'+2 p_0 A_k\right) A_{-k}' -\gamma \frac{F_W}{2} A_k'
A_{-k}' \right]   . \label{consistent}
\ee
After appending this necessary term to the action
Eq.(\ref{AxAction}), we may then write the action as
\be
S_{\textrm{total}} =- \frac{N^2 r_0^2}{16 \pi^2 R^4}
\int \frac{d^4k}{(2 \pi)^4} \int_0^1 du \,
\left[ \frac{1}{2}A_{-k} {\mathcal{L}} \, A_k
+\,\partial_u \Phi \right]  \, . \label{ActionOnShell}
\ee
where ${\mathcal{L}} A_k =0$ is simply the equation of motion of
Eq.(\ref{FullEOM}), and $\Phi$ is a boundary term. Upon evaluating
the on-shell action, the only surviving term is the boundary term,
as we expect from holography. This is given by
\ba
\Phi&=&(B-A)A_k'A_{-k}+\frac{1}{2}(C-A')A_k A_{-k} \nonumber \\
&&-E'A_k''A_{-k}+ E A_k''A_{-k}'-E A_k''' A_{-k}
+E \left(p_1 A_k'+2 p_0 A_k\right) A_{-k}'-\frac{F'}{2} A_k' A_{-k} \, .\label{finalb}
\ea
Therefore, the strategy is to solve the equation of motion
Eq.(\ref{FullEOM}), then insert the solution into the action, which
leaves us only with the boundary term $\Phi$. This is what was done
in the main text of the paper, with massive simplifications arising
from the fact that the only term which contributes to the boundary
action is $B_1$ from the first term of Eq.(\ref{finalb}). This is
because the coefficients $A_W,\,B_W,\,C_W,\,E_W$ and $F_W$ all have
high positive powers of $u$, which, coupled with the regularity of
the solutions of $A_k(u)$ at $u=0$, means that their contribution to
the boundary term vanishes. Let us now list the various coefficients
used in this section, for the contribution of $C^4$ calculated in
the previous appendix. First we have the coefficients with no
$\gamma$-dependence $p_0$ and $p_1$, given by
\be
p_0= \frac{\varpi_0^2-f(u)\kappa_0^2}{u f^2(u)} \quad \textrm{and} \quad p_1=\frac{f'(u)}{f(u)} \, ,
\ee
where $\varpi_0=\omega/(2 \pi T)$ and $\kappa_0=q/(2 \pi T)$. For the
coefficients originating from the $F^2$ term in the action of the
gauge field, we obtain
\ba
B_1&=&\frac{K(u)f(u) L^7(u)}{P(u)} \, , \nonumber \\
D_1&=&-K(u)P(u) L^7(u)\left[\frac{\varpi^2-f(u)K^2(u)\kappa^2}{u f(u)K^2(u)} \right] \, ,
\ea
where $\varpi=\omega R^2/(2 r_0)$ and  $\kappa=q R^2/(2 r_0)$. For the
terms originating from the higher curvature term in the action, we
have the following expressions, here evaluated only for the $C^4$ operator, 
retaining only the $A_x$ fluctuation:
\ba
A_W&=&-2 u^5\left[15 f(u)\kappa_0^2-17 \varpi_0^2 \right] \, , \nonumber \\
B_W&=&- 4 u^4 \left[(14-67 u^2+78 u^4) - 8 u \varpi_0^2-24 u f(u) \kappa_0^2 \right] \, , \nonumber \\
C_W&=&-4 \frac{u^4}{f(u)} \left[ 3 f(u)(3u^2-8)  \kappa_0^2-41 u^2 \varpi_0^2 \right] \, , \nonumber \\
D_W&=&-\frac{u^3}{ f^2(u)} \left[33 u f^2(u) \kappa_0^4 +33u \varpi_0^4 +126 uf(u)\kappa_0^2 \varpi_0^2 \right. \nonumber \\
&&\left. +4 f^2(u)(7 u^2-20)\kappa_0^2+4 (16-45 u^2 +2 u^4)\varpi_0^2\right] \, ,\nonumber \\
E_W&=& -33 u^6 f^2(u)  \, , \nonumber \\
F_W&=& 4 u^5 f(u) (53 u^2-28) \, .
\ea
We remind the reader that although we have computed these
coefficients explicitly only for the $C^4$ operator (see appendix
A), the full set of 10D eight-derivative terms are expected to
{\emph{have the same leading power dependence in $u$}}, as explained
in section \ref{10Dcorrections}. To demonstrate that none of these
coefficients contribute to the on-shell action, we write
$A_k=A_{0}+\gamma A_{1}$, in order to solve the equations of motion
perturbatively in $\gamma$. Now, the equation of motion for $A_{0}$
is then simply given by
\be
A_{0}''+p_1 A_{0}'+ p_0 A_{0} \, = 0 \, .
\ee
This equation has a regular singular point at $u=0$, with indices
$\sigma=0,1$. Expanding around the point $u=0$, we can therefore
write a general solution of the equation in the form $A_{0} = a+ b u
+ c u \log(u)+\cdots$. Now, because we are only interested in the
on-shell action to ${\mathcal{O}}(\gamma)$, and given that
$A_W,B_W,C_W,D_W,E_W,F_W$ in Eq.(\ref{finalb}) all start at
${\mathcal{O}}(\gamma)$, it is clear that the only term contributing
to the on-shell action is $B_1 \, A_k' \, A_{-k}$. We must however
also show that the solution of the equation of motion of $A_{1}$ in the 
ultraviolet region is unaffected by the terms $A_W,B_W,C_W,D_W,E_W,F_W$. Plugging the ansatz
$A_k=A_{0}+\gamma A_{1}$ into the EOM of Eq.(\ref{FullEOM}), we get
the equation of motion of $A_{1}$ as
\be
\partial_u \left[ 2  B_1|_{\gamma\to 0} A_{1}'\right]-2  D_1|_{\gamma\to 0} A_{1} \, = V(A_{0}) \, .
\ee
where
\ba
&&A_W \, A_0''+C_W A_0'+2 \left((D_1-D_1|_{\gamma\to 0})+D_W\right) A_0 \nonumber \\
&&-\partial_u\left(2 (B_1-B_1|_{\gamma\to 0}) A_0'+2 B_W A_0'+C_W A_0 +F_W A_0'' \right) \nonumber \\
&&+\partial_u^2 \left(A_W A_0+2 E_W A_0''+F_W A_0'\right)\, = V(A_{0}) \, .
\ea
We are only interested in the solution of $A_1$ in the region of
small $u$. Taking into account the power dependence of the
coefficients $A_W,B_W,C_W,D_W,E_W,F_W$ and the behaviour of $A_0$
near the boundary, we find that the only contributing factors in the
potential $V(A_0)$ are those given by $2 \left((D_1-D_1|_{\gamma\to
0})\right) A_0 -\partial_u\left(2 (B_1-B_1|_{\gamma\to 0}\right)
A_0'$. We have therefore showed that the only relevant terms are
$B_1$ and $D_1$, both of which arise from the minimal kinetic
operator $F^2$. This completes the analysis for $A_x$. The exercise is
entirely analogous for the other fluctuations.

\section{Appendix: Scattering at low energy and finite 't Hooft coupling}

In this appendix we perform the analysis of the EOMs for low
energies, $\kappa \ll {\mathcal{K}}^3$, which is equivalent to
considering the low temperature regime $q T^2 \ll Q^3$ when the
virtuality $Q^2$ is fixed. For the small $u$ region  $0 \leq u \leq
1/{\mathcal{K}}_0^2 \ll 1$, together with the assumption that $A_i
\, \kappa_0^2 \gg 2 \, A_i'$ and similarly for $a$, in the limit
$\lambda\rightarrow\infty$, equations (\ref{Aicorrected}) and
(\ref{acorrected}) become
\be
A_i'' - \frac{{\mathcal{K}}_0^2}{u} \, A_i  = - \, \kappa_0^2 \, u \,
A_i  \, ,
\ee
and
\be
a'' + \frac{1}{u} a' - \frac{{\mathcal{K}}_0^2}{u} \, a  = -
\kappa_0^2 \, u \, a \, .
\ee
Now, consider the case when the 't Hooft coupling is finite. Within
the same level of approximation as before, and within the parametric
region $0 \leq u \leq 1/{\mathcal{K}}^2 \ll 1$, equations
(\ref{Aicorrected}) and (\ref{acorrected}) become
\be
A_i'' - \frac{{\mathcal{K}}^2}{u} A_i = -  \left(1 + \frac{325}{4} \,
\gamma \right) \, \kappa^2 \, u \, A_i \, , \label{AilowEe}
\ee
and
\ba
a'' + \frac{1}{u} \, a' - \frac{{\mathcal{K}}^2}{u} \, a = - \left(1
+ \frac{325}{4} \, \gamma  \right) \, \kappa^2 \, u \, a \, .
\label{alowEe}
\ea
The next step is to solve these equations in perturbation theory.
For that purpose it is convenient to define $\xi = 2 \,
{\mathcal{K}} \, u^{1/2}$, so that the equations (\ref{AilowEe}) and
(\ref{alowEe}) become
\be
\left(\frac{d^2}{d\xi^2} + \frac{1}{\xi} \, \frac{d}{d\xi} - 1 -
\frac{1}{\xi^2} \right) \, h(\xi) = - \left(1 + \frac{325}{4} \,
\gamma \right) \, \left(\frac{ \kappa^2 \, \xi^4}{16 \,
{\mathcal{K}}^6} \right) \, h(\xi) \, , \label{hlowE}
\ee
and
\be
\left(\frac{d^2}{d\xi^2} + \frac{1}{\xi} \, \frac{d}{d\xi} - 1
\right) \, a(\xi) = - \left(1 + \frac{325}{4} \, \gamma \right) \,
\left(\frac{ \kappa^2 \, \xi^4}{16 \, {\mathcal{K}}^6} \right) \,
a(\xi) \, , \label{aalowE}
\ee
respectively, where in equation (\ref{hlowE}) we have redefined
$A_i(\xi) = A_i(0) \, \xi \, h(\xi)$.

In order to study these equations perturbatively, let us begin with
equation (\ref{hlowE}). The zeroth order solution is $h^{(0)}(\xi) =
K_1(\xi)$, where $K_1$ is the modified Bessel function of order one.
The function $\xi \, h^{(0)}(\xi)$ approaches one at the AdS
boundary. The general solution is given as a sum of the general
solution of the homogeneous ordinary differential equation and the
convolution of the Green's function with the source of equation
(\ref{hlowE})
\be
h(\xi) = h^{(0)}(\xi) - \int_0^\infty d\xi' \, G(\xi, \, \xi') \,
\left(1 + \frac{325}{4} \, \gamma \right) \, \left(\frac{ \kappa^2
\, \xi'^4}{16 \, {\mathcal{K}}^6} \right) \, h(\xi') \, ,
\ee
where $G(\xi, \, \xi')$ is the Green's function corresponding to the
differential operator of Eq.(\ref{hlowE}) with appropriate boundary
conditions. This is the same Green's function as given in
\cite{Hatta:2007cs}. The ${\cal {O}}(u)$-perturbation theory
solution for the transverse modes can be expressed as $A_i(u)=
A_i^{(0)}(u) + A_i^{(1)}(u)$, where
\ba
A_i^{(0)}(u) &=& 2 \, {\mathcal{K}} \, A_i(0) \, u^{1/2} \, K_1[2 \, {\mathcal{K}} \, u^{1/2}] \, , \\
A_i^{(1)}(u) &=&  A_i(0) \, \frac{\kappa^2}{5 \, {\mathcal{K}}^4} \,
\left(1 + \frac{325}{4} \, \gamma \right) \, u \, .
\ea
For small values of $u$, we can expand the modified Bessel function.
Thus we obtain
\be
A_i^{(0)}(u) \simeq  A_i(0) \, (1 + u \, {\mathcal{K}}^2 \, [\log
{\mathcal{K}}^2 - 1 + \log u + 2 \, \xi_E]) \, ,
\ee
where $\xi_E$ is the Euler-Gamma function which is $0.57 \cdot \cdot
\cdot$.

The solution for the longitudinal modes in perturbation theory can
be expressed as $a(u)= a^{(0)}(u) + a^{(1)}(u)$, where
\ba
a^{(0)}(u) &=& - 2 \, \kappa^2 \, A_L(0) \, K_0[2 \, {\mathcal{K}} \, u^{1/2}] \, , \\
a^{(1)}(u) &=& - 2 \, A_L(0) \, \frac{\kappa^4}{15 \,
{\mathcal{K}}^6} \, \left(1 + \frac{325}{4} \, \gamma \right) \, .
\ea
For small values of $u$ we can expand the modified Bessel function as follows
\be
a^{(0)}(u) \simeq \kappa^2 \,  A_L(0) \, [\log {\mathcal{K}}^2 +
\log u + 2 \, \xi_E] \, .
\ee
Inserting these approximations for the transverse and longitudinal
components of the gauge fields into the on-shell action, we obtain
the results for the low energy scattering presented in section 5.1.

\end{document}